\documentclass[fleqn,usenatbib]{mnras}

\usepackage{newtxtext,amsmath, xcolor}
\usepackage[flushleft]{threeparttable}

\usepackage[T1]{fontenc}

\DeclareRobustCommand{\VAN}[3]{#2}
\let\VANthebibliography\thebibliography
\def\thebibliography{\DeclareRobustCommand{\VAN}[3]{##3}\VANthebibliography}

\usepackage{graphicx}	
\usepackage{amsmath}	
\usepackage{amssymb}	
\usepackage{nicefrac} 
\usepackage{xfrac}    
\usepackage{hyperref}

\usepackage[normalem]{ulem} 

\usepackage{fontawesome5}

\newcommand{\rockstar}{\textsc{Rockstar}}
\newcommand{\ctrees}{\textsc{Consistent Trees}}
\newcommand{\hmf}{\textsc{HMF}}
\newcommand{\dlogndz}{$d \log n_\text{vir} /dz \sim 0.2$}
\newcommand{\msmvir}{$M_\ast-M_\text{vir}$}
\newcommand{\msteobs}{$M_{\text{obs},\ast}$}
\newcommand{\mvirthreetimestwelve}{$M_\text{vir} \sim 3\times 10^{12}\,M_\odot$}
\newcommand{\Mbar}{$f_\text{bar} \; M_\text{vir}$}

\title[Galaxy Number Densities \& Evolution]{On matching galaxy number densities to reconstruct galaxy evolutionary tracks}

\author[Aldo Rodriguez-Puebla et al.]{
Aldo Rodr\'{i}guez-Puebla,$^{1}$\thanks{E-mail: apuebla@astro.unam.mx}
Vladimir Avila-Reese,$^{1}$
Joel R. Primack$^{2}$ and 
Carlo Cannarozzo$^{3,4}$ 
\\
$^{1}$Universidad Nacional Aut\'onoma de M\'exico, Instituto de Astronom\'ia, A. P. 70-264, 04510, Ciudad de M\'exico, M\'exico\\
$^{2}$Department of Physics, University of California, Santa Cruz, CA 95064, USA\\
$^{3}$New York University Abu Dhabi, PO Box 129188, Abu Dhabi, United Arab Emirates \\
$^{4}$Center for Astrophysics and Space Science (CASS), New York University Abu Dhabi\\ 
}

\date{Accepted XXX. Received YYY; in original form ZZZ}

\pubyear{\the\year{}}

\begin{document}
\label{firstpage}
\pagerange{\pageref{firstpage}--\pageref{lastpage}}
\maketitle

\begin{abstract}
The cumulative number density matching approach equates number densities between adjacent redshifts to derive empirical galaxy evolution tracks from the observed galaxy stellar mass function. However, it is well known that this approach overlooks scatter in mass assembly histories and merger effects, with previous studies relying on model-based corrections, either from hydrodynamical cosmological simulations oradjustments to the evolution of cumulative number density with redshift.  Here, we revisit this approach, showing that dark matter halo assembly histories imply evolving number densities that are far from constant. These exhibit an average slope of \dlogndz\ dex for progenitors at $z=0$, leading to evolutionary tracks where galaxies are $\sim2-3$ times smaller in mass at $z\sim2$ and an order of magnitude smaller by $z\sim7$ compared to the number density matching approach. We show that evolving halo number densities provide realistic evolutionary tracks without relying on model-based corrections. Accounting for random errors in stellar mass measurements is also crucial for robust track derivation. We also discuss a generalization that incorporates a galaxy’s star formation activity. When additionally considering the scatter around the \msmvir\ relation ($\sim0.15$ dex), our evolving halo cumulative number density approach shows that some observed stellar masses, \msteobs, can exceed the universal baryon fraction $f_\text{bar}\sim0.16$. For instance, at $z=5$, around $2\%$ of progenitor galaxies of haloes with \mvirthreetimestwelve\ have \msteobs>\Mbar, suggesting a potential ``early galaxy formation problem''. However, when deconvolving mass from random errors this tension is reduced with significant confidence at the $\sim5-6\sigma$ level.

\end{abstract}

\begin{keywords}
galaxies: luminosity function, mass function -- galaxies: evolution -- galaxies: haloes -- methods: statistical -- cosmology: dark matter.
\end{keywords}



\section{Introduction}

The galaxy stellar mass function (GSMF) is one of the most important tools in extragalactic astronomy: it describes the abundance of galaxies per unit of comoving volume as a function of their stellar mass at a given redshift. Moreover, since the GSMF is a time-integrated distribution, it reflects the cumulative processes related to stellar mass growth through in situ star formation and galaxy mergers \citep[e.g.,][]{DroryAlvarez2008}. The characteristic Schechter-like shape of the GSMF, with its steepening at the low-mass end, flattening for intermediate-mass galaxies, and sharp drop-off at the high-mass end, provides crucial insights into the heating/cooling processes of the gaseous haloes, star formation from cold infalling gas, merging of galaxies, and the supernova and active galactic nuclei (AGN) feedback processes that regulate the baryonic cycle and drive to quenching star formation in galaxies \citep[see e.g.,][]{Peng+2010,Somerville+2015,Primack2024}. Thanks to the unprecedented number of galaxy surveys spanning a wide range of masses and redshifts, we can now trace the evolution of the GSMF to very high redshifts \citep[see e.g.,][]{Conselice+2016,Weaver+2023,Harvey+2024,Weibel+2024} providing insights into the main processes governing galaxy evolution.

By comparing the GSMF at different redshifts, we can derive evolutionary tracks that show how galaxies grow in mass over time \citep{DroryAlvarez2008,Peng+2010}. One of the most popular and straightforward approaches is to match cumulative number densities\footnote{If the GSMF is defined as the number of galaxies within the mass range $\log M_\ast \pm d \log M_\ast /2$ per comoving volume per dex, $\phi_\ast(M_\ast)$, then the cumulative number density is given by \begin{equation*} n_{\text{gal}}(>M_\ast) = \int_{M_\ast}^\infty \phi_\ast(M) d\log M. \end{equation*} } between adjacent redshifts \citep{LoebPeebles2003,vanDokkum+2010,Patel+2013,Leja+2013,Behroozi+2013f,Clauwens+2016,vandeVoort_2016,Davidzon+2017,Wang+2023}. However, this method has limitations, as it ignores the effects of mergers and the scatter in the mass assembly histories of galaxies, as noted by previous authors \citep[e.g.,][]{Behroozi+2013f,Clauwens+2016,Wang+2023}. Semi-empirical modeling of the galaxy-halo connection, which traces galaxy mergers and star formation histories \citep[][see \citealp{Wechsler+2018} for a review and more references therein]{Conroy+2009,Firmani+2010a,Rodriguez-Puebla+2017,Moster+2018,Behroozi+2019,Grylls+2020,Fu+2024}, is a valuable approach to addressing these limitations.

Indeed, previous studies have used semi-empirical approach of the galaxy-halo connection to account for scatter in galaxy progenitors and mergers in the cumulative number density approach. For example, \citet{Behroozi+2013f} used (sub)halo abundance matching (SHAM) and dark matter halo assembly histories to estimate the median change in cumulative number densities, finding that this change is well approximated by an increase of $\sim0.16$ dex per $\Delta z$ for all galaxies with final masses defined at $z=0$. Here we explore this assumption and generalize the study to the progenitors of galaxies defined at different redshifts and not only at $z=0$. Alternatively, some studies have proposed using hydrodynamical cosmological simulations to address these issues \citep[see e.g.,][]{Clauwens+2016,vandeVoort_2016,Wang+2023}. In particular, using the EAGLE simulation \citep{Crain+2015}, \citet{Clauwens+2016} investigated the cumulative number density evolution of star-forming and quiescent galaxies, showing that the number density of progenitors of galaxies of masses $M_\ast = 10^{10.75}M_\odot$ strongly depends on the galaxies' star formation activity. Nonetheless, the authors found that when considering all galaxies together (star-forming and quiescent), the \citet{Behroozi+2013f} prescription accurately predicts the evolution of the median cumulative number density.

Another relevant study is that of \citet{Wang+2023}, who developed a statistical framework for linking galaxies with masses $M_\ast \gtrsim10^{10}M_\odot$ to dark matter haloes at different redshifts to infer the evolution of the galaxy population. Unlike previous methods that rely solely on stellar mass, their framework incorporates the star formation state of descendant galaxies. Specifically, they introduce a conditional age distribution matching method \citep{Hearin_Watson2013, Masaki+2013, Watson15} to link star formation activity to halo formation properties, refining the galaxy–halo connection. However, note that \citet{Kakos+2024} found that such models are inconsistent when compared to two-point cross-correlations. These authors found by testing their approach in both the EAGLE and Illustris \citep{Pillepich+2018} hydrodynamic cosmological simulations that these tend to over estimate the progenitor mass of the galaxies compared to what they inferred from observations. 

While the approaches employed by \citet{Clauwens+2016} and, more recently, \citet{Wang+2023} seem to offer a more general framework than that of \citet{Behroozi+2013f}, they depart from the original intent of cumulative number density matching; which was to provide a simple but effective statistical tool for tracking galaxy evolution over time In addition, these methods are constrained by the resolution of the simulations, limited in the mass range, and focused on studying the progenitors of galaxies with final masses defined only at $z=0$.

With the upcoming generation of large galaxy surveys, it is crucial to develop fast, simple and, more importantly, robust tools that can inform us about galaxy evolution, while \emph{minimizing} assumptions. In this paper, we present an alternative approach that is based on the assumption that galaxies form and evolve within dark matter haloes, resulting in a monotonic relationship between stellar and halo masses, with some associated scatter. Similar to \citet{Behroozi+2013f}, in this paper we employ the median aggregation (or accretion) histories (MAHs) of dark matter haloes from $N$-body cosmological simulations, we infer evolutionary tracks of halo cumulative number densities, which can then be matched to galaxies using the (sub)halo abundance matching approach \citep{ValeOstriker04,Conroy+2006}, after accounting for random errors in the intrinsic scatter of the stellar-to-halo mass relation. By construction, these tracks account for the effects of scatter in mass assembly histories and mergers. We demonstrate that the constant increase of  $\sim0.16$ dex per $\Delta z$ proposed by \citet{Behroozi+2013f} is more effective for intermediate halo/galaxy masses and low redshifts. Besides, this slope likely depends on cosmological parameters. In contrast, the (semi)empirical approach discussed in this paper is  straightforward and produces realistic evolutionary tracks for galaxies. Additionally, we discuss how this approach can be easily generalized to galaxies with arbitrary star formation activity, resulting in evolving galaxy number densities that align with previous findings based on more complex models \citep{Wang+2023}\footnote{A public version of \textsc{MatchA}, the Matching Abundances code, is available at this  \href{https://github.com/TheConCHaProject/ConCHa}{GitHub link \faGithub}.}.

Finally, as an immediate application, we apply this method to explore the "impossible galaxies" problem in light of the recent discovery of massive galaxies that {apparently are too massive to be explained using simple halo-based arguments at high redshifts \citep[see e.g.,][]{Steinhardt+2016,Labbe+2023,Boylan-Kolchin_2023}. Our results show that once random errors are considered, the early galaxy formation tension is reduced with significant confidence at the $\gtrsim5\sigma$ level. 

This paper is organized as follows. In Section \ref{sec:random_errors} we discuss the random errors in stellar mass estimates and their impact on the GSMF. In Section \ref{sec:num_match}  we describe the cumulative number density matching approach and compare results when using the true GSMF versus the observed one. Section \ref{sec:evol_halo_nvir} focuses on deriving halo assembly histories, demonstrating the strong evolution of halo cumulative number densities with redshift. These results are then used in Section \ref{sec:proj_halo_den_onto_gal_den} to project halo number densities onto galaxy number densities and, ultimately, galaxy evolutionary tracks. In Section \ref{sec:discussion} we discuss implications for the galaxy-halo connection, a generalization to our model to account for the star formation activity  of the galaxy and end that section by presenting our new semi-empirical model in the context of the ``impossible galaxies'' problem. Finally, in Section \ref{sec:conclusions} provides our conclusions.

In this paper we assume the following cosmology: $\Omega_\text{m} = 0.307$, $\Omega_\Lambda = 1-\Omega_\text{m}$, $\Omega_\text{b} = 0.048$, $n_s=0.96$, $h=0.678$, and $\sigma_8 = 0.823$, which is close to the \cite{Planck+2015} cosmology and the Bolshoi-Planck simulation \citep{Klypin+2016,Rodriguez-Puebla+2016}. Additionally, we assume the initial mass function of \cite{Chabrier2003}.

\section{The observed GSMF is affected by random errors}
\label{sec:random_errors}

Stellar mass estimates of individual galaxies are subject to random errors \citep[for a disussion see,][]{Conroy2013}. Therefore, the observed GSMF is the result of convolving the true GSMF with the random errors \citep{Cattaneo+2008,Behroozi+2010,Ilbert+2013,Grazian+2015,Rodriguez-Puebla+2017,Rodriguez-Puebla+2020}:
\begin{equation}
    \phi_{\ast,\text{conv}}(M_\ast) = \int_{-\infty}^{\infty} \mathcal{P}(x-\log M_{\ast}) \, \phi_{\ast,\text{true}}(x) d x,
    \label{eq:gsmf}
\end{equation}
where $\mathcal{P}(x)$ is the probability distribution of errors and $\phi_\ast$ is in units of Mpc$^{-3}$ dex$^{-1}$. The functional form of $\mathcal{P}(x)$ is sometimes assumed to be a Gaussian distribution, $\mathcal{G}(x)$, but also as a product of a Gaussian and Lorentzian distribution, $\mathcal{L}(x)$, \citep[][see \citealp{Leja+2020} for a similar result]{Ilbert+2013,Davidzon+2017}. For simplicity, in this paper we assume that $\mathcal{P} = \mathcal{G}$ is a Gaussian distribution. 

As discussed by previous authors \citep{Behroozi+2010,Cattaneo+2008,Wetzel+2010,Behroozi+2013,Moster+2013,Rodriguez-Puebla+2017} Eq. (\ref{eq:gsmf}) implies that the observed count of galaxies at a given stellar mass deviates from the true count due to the \citet{Eddington1913,Eddington1940} bias. Specifically, random errors tend to upscatter many more low-mass galaxies into higher mass bins because of the steep drop-off in the GSMF with increasing $M_\ast$. 

To gain insights on the general effect of random errors over the true GSMF, we can use a crude approximation to the convolved GSMF (Rodriguez-Puebla, in prep.):
\begin{equation}
    \phi_{\ast,\text{conv}}(M_\ast) \approx \phi_{\ast,\text{true}}(M_\ast) + 
    \phi_{\ast,\text{true}}''(M_\ast) \frac{\sigma_{\ast,\text{ran}}^2}{2},
    \label{eq:convGSMF_approx}
\end{equation}
where the $\phi_{\ast,\text{true}}''$ indicates the second derivative with respect to $\log M_\ast$ and $\sigma_{\ast,\text{ran}}$ is the variance of the Gaussian distribution. The above is derived under the assumption that random errors are Gaussian-distributed and that $\mathcal{G}$ is negligible outside a certain boundary around the value of $\log M_\ast$, typically around $3-5\sigma_{\ast,\text{ran}}$. By expanding in a Taylor series around $M_\ast$, it is easy to show that Eq. (\ref{eq:convGSMF_approx}) is valid. Qualitatively, the true (or deconvolved) GSMF is expected to be smaller than the convolved one since the second term in Eq. (\ref{eq:convGSMF_approx}) is always positive for a Schechter function. 

Assuming that the GSMF is well described by a Schechter function, at lower masses the GSMF is given by $\phi_{\ast,\text{true}} \propto M_{\ast}^{\alpha+1}$. Using the value of $\alpha = -1.4$ from \cite{Peng+2010}, and assuming a representative error value of $\sigma_{\ast,\text{ran}}\sim 0.2$ dex, see Figure \ref{fig:ran_ms} described below, the correction to the GSMF at lower masses is $\sim 1.6\%$. In contrast, at higher masses where the GSMF decreases exponentially as $\phi_{\ast,\text{true}} \propto \exp(-M_\ast/M_{c})$, where $M_c$ is the Schechter characteristic mass and for the most massive galaxies with $M_\ast/M_{c} \gg 1$, the convolved GSMF is given by:
\begin{equation}
    \phi_{\ast,\text{conv}}(M_\ast) \approx  \phi_{\ast,\text{true}}(M_\ast) \left[ 1 + \frac{\sigma_{\ast,\text{ran}}^2}{2} \left(\frac{M_\ast}{M_{c}}  \right)^2   \right].
\end{equation}
That is, the convolved GSMF increases relative to the true GSMF as the square of the mass. Recall that Eq. (\ref{eq:convGSMF_approx}) provides only a crude approximation to understand the effect of random errors. Later in this paper, we will present a more accurate assessment of the convolution effect. Next, we discuss observational studies that have constrained the value of random errors as a function of redshift.

\begin{figure}
		\includegraphics[width=\columnwidth]{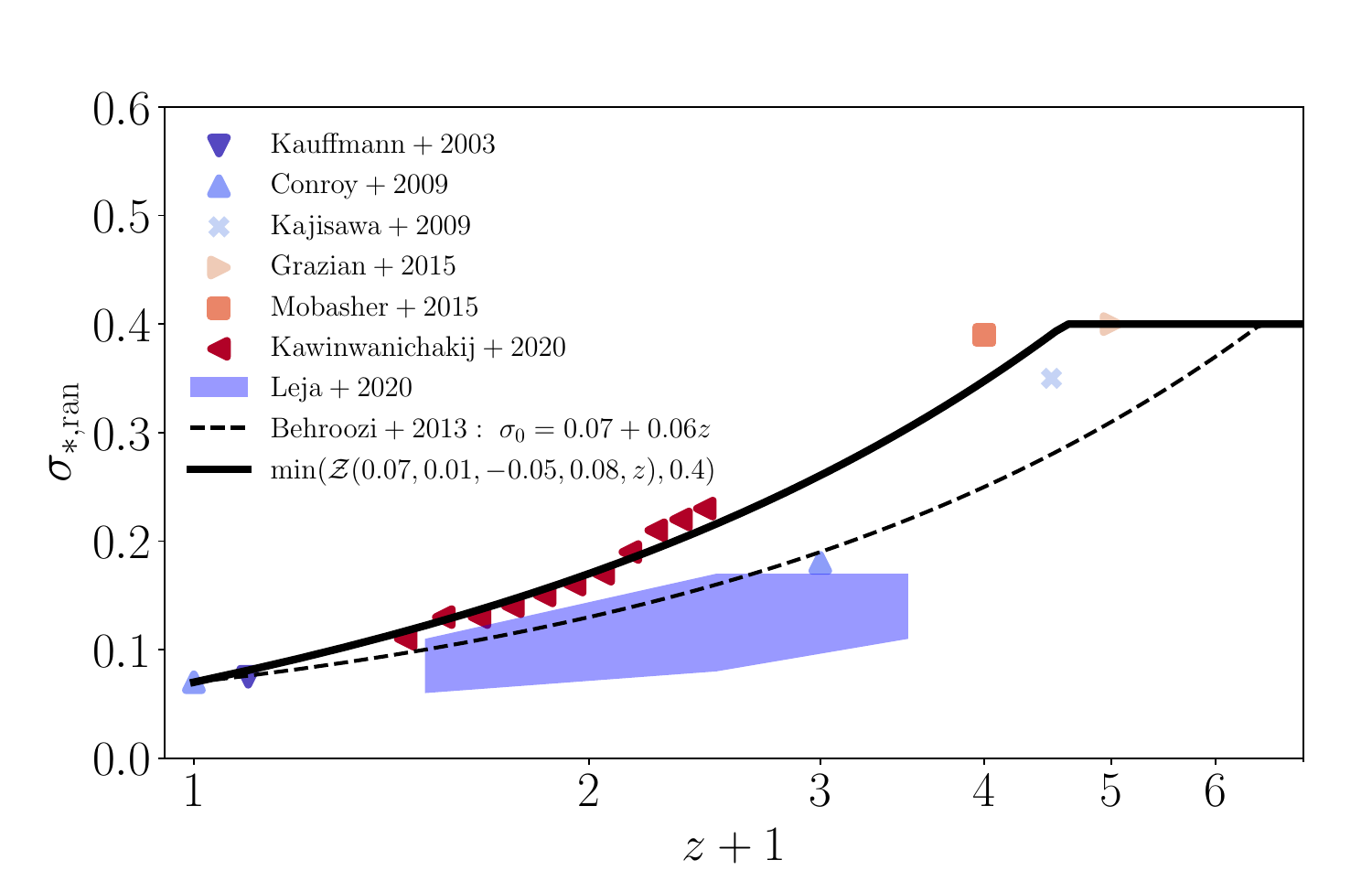}
			\caption{Redshift evolution of random errors in stellar masses. Errors range from $\sim0.1$ dex to $\sim0.4$ dex at higher redshifts. The solid line shows the model that best describes the data, Eq. (\ref{eq:ran_error}), while the dashed line is the model proposed by \citet{Behroozi+2013}.
 		}
	\label{fig:ran_ms}
\end{figure}

Figure \ref{fig:ran_ms} shows how the $\sigma_{\ast,\text{ran}}$ changes with redshift from different works \citep[][]{Kauffmann+2003,Conroy+2009a,Kajisawa+2009,Grazian+2015,Mobasher+2015,Kawinwanichakij+2021,Leja+2020}.\footnote{In the case of \citet{Leja+2020}, their errors depend not only on redshift $z$ but also on mass. We show the range of their errors as shaded areas, representing the upper and lower values for a given $z$.} The first observation from this figure is that random errors increase with redshift. Errors range from $\sim0.1$ dex at low redshifts to around $\sim0.4$ dex at higher redshifts. In this paper, we adopt the model by Rodriguez-Puebla et al. (in prep.)  to describe most of the data in Figure \ref{fig:ran_ms} given by:
\begin{equation}
    \sigma_{\ast, \text{ran}} = \text{min} \left( \mathcal{Z} (0.07,0.01,-0.05,0.08,z), 0.4 \right),
    \label{eq:error_evol_fiducial}
\end{equation}
where
\begin{equation}
	\mathcal{Z} (p_1,p_2,p_3, p_4,z)  = p_1+p_2(1-a) + p_3 \log a + p_4 z,
    \label{eq:ran_error}
\end{equation}
with $a = (1+z)^{-1}$ representing the scale factor. This model of random errors is represented by the solid black line in Figure \ref{fig:ran_ms}. For comparison we reproduce the model derived by \cite{Behroozi+2013} which leads to smaller errors for a given stellar mass. The discussion of random errors will be detailed in a future paper. For the scope of this paper, we use a more conservative model for the upper values of $\sigma_{\ast,\text{ran}}$, as given by Eq. \ref{eq:error_evol_fiducial}.

Equation (\ref{eq:gsmf}) is assumed to be valid for a given $z$. Note, however, that observational determinations are made in redshift intervals. Over a redshift interval between $z$ and $z + \Delta z$ the resulting GSMF is the volume-weighted average of the evolved GSMF:
\begin{equation}
    \phi_{\ast,\text{obs}}(M_\ast,\Delta z) = \frac{1}{\Delta V_c}\int_{z}^{z+\Delta z} \phi_{\ast,\text{conv}}(M_\ast,z') \, d V_c(z')
    \label{eq:GSMF_obs}
\end{equation}
where $V_c(z)$ is the comoving volume at $z$ and
\begin{equation}
    \Delta V_c = V_c(z+\Delta z) - V_c(z).
\end{equation}
Hereafter, we will refer to Eq. (\ref{eq:GSMF_obs}) as the observed GSMF. In this paper we are not accounting for redshift errors, but we notice that some of the GSMFs used in \citet{Rodriguez-Puebla+2017} and Rodriguez-Puebla et al. (in prep), which is the data employed here, rely on photometric redshifts. However, we discuss the potential impact of this effect. Including redshift errors would introduce an additional convolution in Eq. (\ref{eq:GSMF_obs}), adding another source of Eddington bias \citep{Stefanon+2015,Leja+2020}, which would further inflate the number density of galaxies.

\begin{figure}
		\includegraphics[width=\columnwidth]{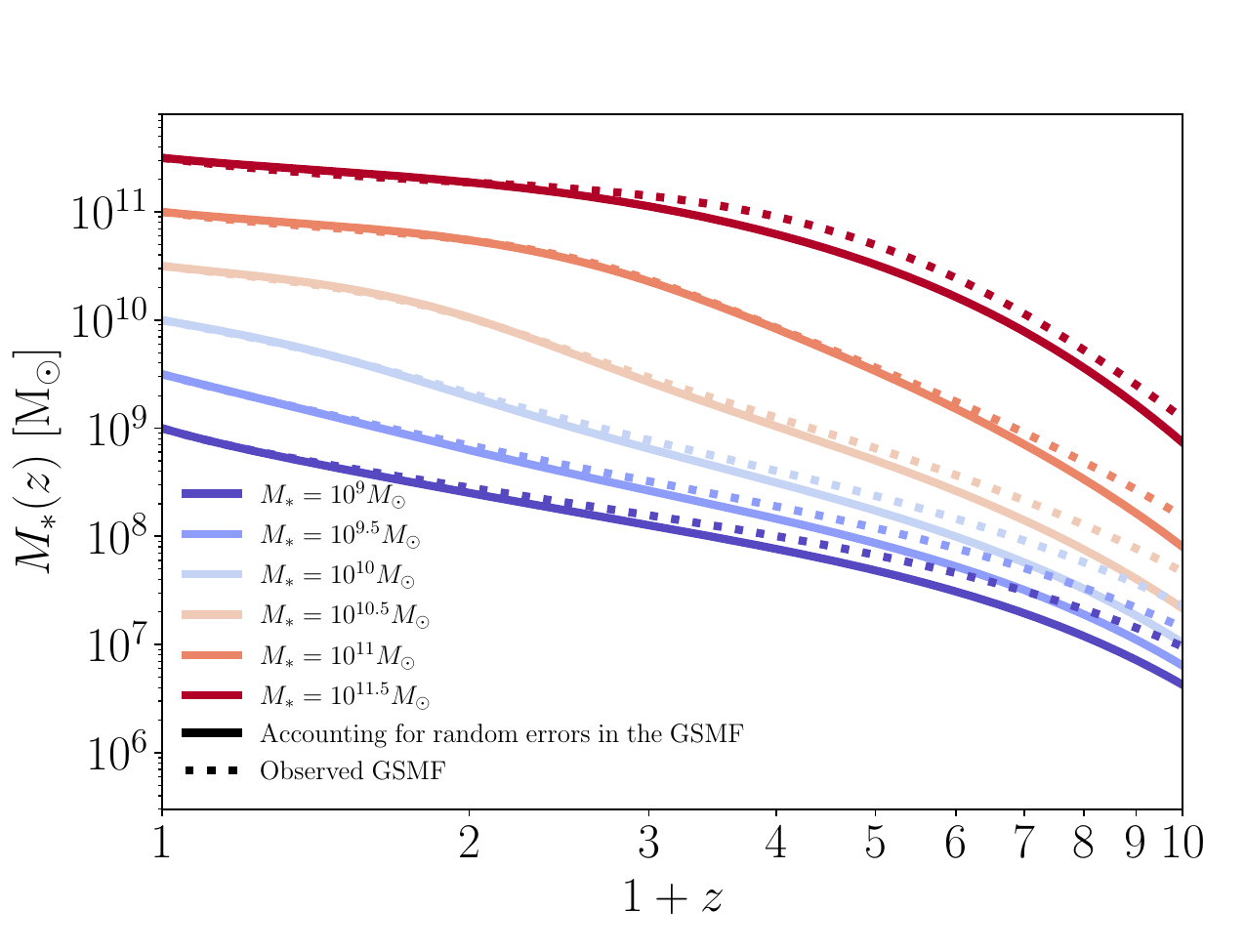}
			\caption{Assembly history of progenitors at $z=0$ for galaxies in the stellar mass range $M_\ast = 10^{9} - 10^{11.5} M_\odot$. The solid lines represent the results of matching cumulative number densities while accounting for random errors, whereas the dotted lines show the results using the observed GSMF as it is. Accounting for random errors leads to smaller stellar masses at a fixed redshift for a given progenitor because of Eddington bias.
 		}
	\label{fig:mass_assembly_obs_vs_deconv}
\end{figure}

\begin{figure}
		\includegraphics[width=\columnwidth]{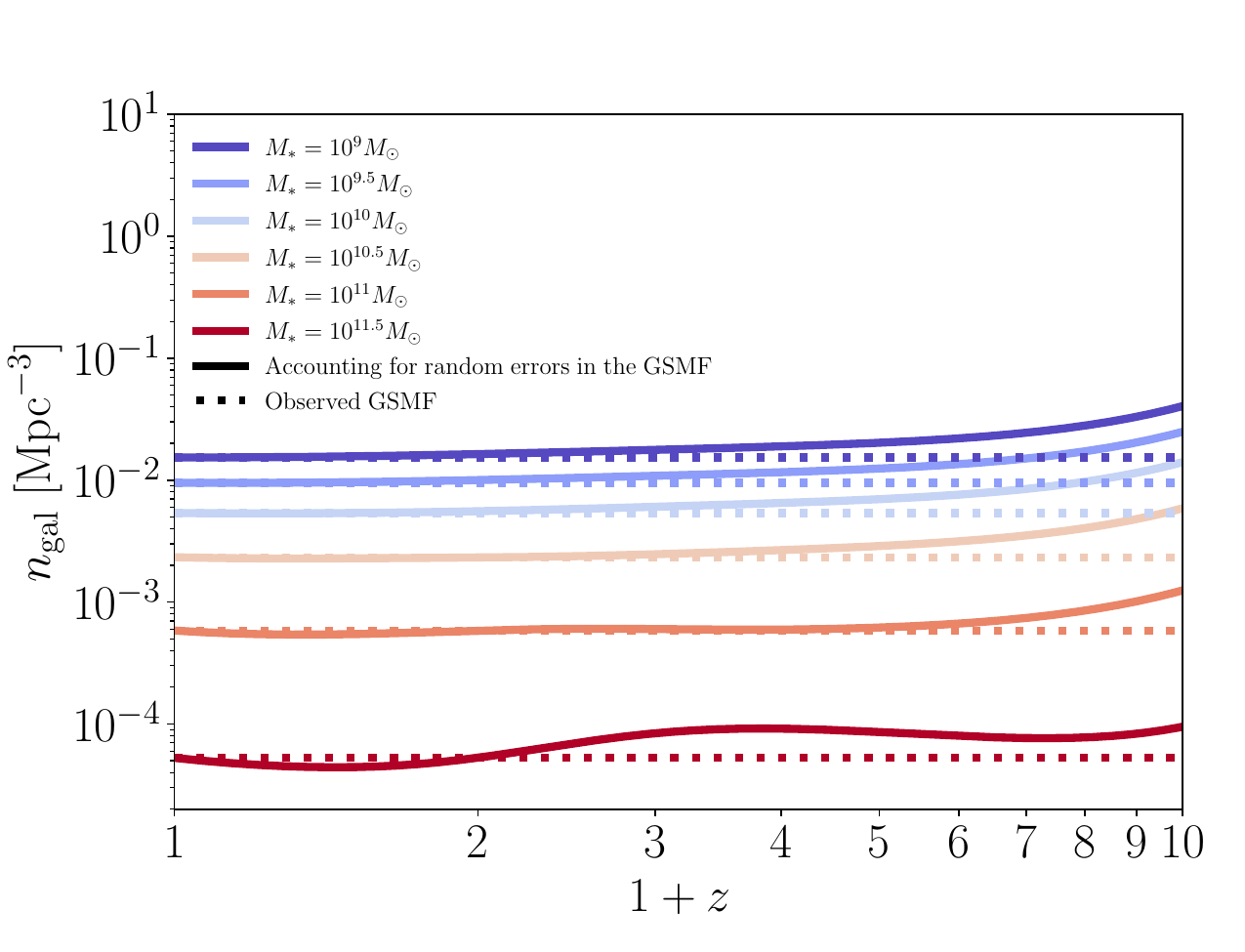}
			\caption{Cumulative number density evolution for the progenitors shown in Fig. \ref{fig:mass_assembly_obs_vs_deconv}. The solid lines represent the observed number density, $n_\text{obs}$, evaluated using tracks that account for random errors in $M_\ast$. The dotted lines indicate constant cumulative number densities for the observed GSMF. This figure highlights the effect of accounting for random errors, which results in increased observed cumulative number densities for all galaxies at high redshift.
         		}
	\label{fig:numb_density_obs_vs_deconv}
\end{figure}

In this paper, we use the best fitting model from Rodriguez-Puebla et al. (in prep., see also figure 5 from \citealp{Rodriguez-Puebla2024}) for both $\phi_{\ast,\text{obs}}$ and $\phi_{\ast,\text{true}}$. Specifically, the authors provide fitted models for both the observed and true GSMFs. To achieve this, the authors compiled and homogenized a large data set of GSMFs from the literature for both all galaxies at $z\sim0-9$, and quiescent galaxies, from $z\sim0-4.5$. This compilation represents an update to the GSMFs presented in \cite{Rodriguez-Puebla+2017} including recent data from the James Webb Space Telescope, JWST \citep[such as][]{Navarro-Carrera+2024,Weibel+2024} and COSMOS2020 \citep{Weaver+2023}, as will be discuss in Rodríguez-Puebla et al. (in prep.). Among the key quantities that the authors homogenized was the Initial Mass Function to that of \citet{Chabrier2003}, and they adopted the same cosmology employed in this work. The authors performed Bayesian fits at the same time to the redshift evolution of the GSMFs simultaneously for all galaxies ($0\leq z \leq 9$) and for quiescent galaxies ($0\leq z \leq 4.5$), assuming that quiescent galaxies are well described by a triple modified-Schechter function (see Vazquez-Mata et al. submitted) and star-forming galaxies by a double modified-Schechter function, see their sections 3.2 and 3.3 for details.

\section{The Empirical approach: Evolutionary tracks by matching cumulative number densities}
\label{sec:num_match}

In this Section, we explore the implications of the redshift evolution of the deconvolved GSMF (hereafter the true GSMF) for galaxy formation and evolution. A common method for deriving \textit{purely empirical} evolutionary tracks based on the observed GSMF is the so-called matching cumulative number density technique \citep{LoebPeebles2003,vanDokkum+2010, Behroozi+2013f}. This technique involves matching cumulative number densities between two adjacent redshifts 
\begin{equation}
    \int_{M_\ast}^{\infty} \phi_{\ast}(x,z) dx =
    \int_{M_\ast'}^{\infty} \phi_{\ast}(x,z+\Delta z) dx, 
    \label{eq:num_den}
\end{equation}
where $M_\ast(z)>M_\ast'(z+\Delta z)$ such that $M_\ast'(z+\Delta z) = M_\ast(z) - \Delta M_\ast(\Delta z)$. Equation (\ref{eq:num_den}) is essentially the integral form of the continuity equation for $\phi_{\ast}(M_\ast, z)$, which describes changes in the GSMF due to star formation \citep[see equation 1 in][]{DroryAlvarez2008}, in the limit as $\Delta z \rightarrow 0$. To illustrate this, we assume, without loss of generality, that $\Phi_\ast$ is the GSMF in units of Mpc${^{-3}}$ $M_\odot^{-1}$,\footnote{Note that $\phi_\ast$ and $\Phi$ are related by $\Phi(M_\ast) = \phi(M_\ast) \cdot M_\ast \cdot \ln 10$.} and under this assumption, we can rewrite Eq. (\ref{eq:num_den}) as follows\footnote{Notice that in both Eqs. (\ref{eq:num_den}) and (\ref{eq:new_var_match}) the variable $x$ is context-dependent, and its meaning and units change accordingly. }
\begin{align}
    \int_{M_\ast}^{\infty} \Phi_{\ast}(x,t) dx - & \int_{M_\ast'}^{\infty} \Phi_{\ast}(x,t) dx =  \,  \nonumber \\
    & \int_{M_\ast'}^{\infty} \Phi_{\ast}(x,t+\delta t) dx - \int_{M_\ast'}^{\infty} \Phi_{\ast}(x,t) dx, 
    \label{eq:new_var_match}
\end{align}
where $t$ is a look-back time evaluated at $z$ and $\delta t$ corresponds to the increase in redshift when $\Delta z\rightarrow0$ and $M'_\ast(t+\delta t) = M_\ast(t) - \delta M_\ast(\delta t)$. The above eq. can be rewritten as follows:
\begin{equation}
    -\int_{M_\ast'}^{M_\ast} \Phi_{\ast}(x,t) dx = \int_{M_\ast'}^{\infty} \left[ \Phi_{\ast}(x,t+\delta t) - \Phi_{\ast}(x,t) \right] dx.
\end{equation}
If $\delta t$ is small then 
\begin{equation}
    -\Phi_{\ast}(M_\ast,t)\delta M_\ast = \int_{M_\ast'}^{\infty} \frac{\partial \Phi_\ast(x,t)}{\partial t} \delta t dx,
\end{equation}
or 
\begin{equation}
    -\Phi_{\ast}(M_\ast,t) \dot{M}_\ast = \int_{M_\ast'}^{\infty} \frac{\partial \Phi_\ast(x,t)}{\partial t} dx.
\end{equation}
Finally, the above equation reduces to the continuity equation in $\Phi_\ast$ as
\begin{equation}
    -\frac{\partial}{\partial M_\ast}  \left[\Phi_{\ast}(M_\ast,t) \dot{M}_\ast\right] = \frac{\partial \Phi_\ast(M_\ast,t)}{\partial t}.
    \label{eq:continuity_eq}
\end{equation}

As shown by the continuity equation (Eq. \ref{eq:continuity_eq}) this approach has limitations, as it overlooks scatter in the mass assembly histories of galaxies and the effects of mergers. To obtain more realistic evolutionary tracks, corrections based on previous galaxy formation models should be incorporated \citep[see, e.g.,][]{Behroozi+2013f, Clauwens+2016,vandeVoort_2016,Wang+2023}. Another drawback is that authors typically use the \emph{observed} GSMFs. This is wrong since as discussed earlier in this section, the observed GSMF is the result of the true GSMF convolved with random errors, and inferences will thus carry selection biases due to \citet{Eddington1913,Eddington1940} bias. Therefore, in this paper we propose that the next logical improvement in this technique is the use of the more appropriate true GSMF:
\begin{equation}
    \phi_{\ast} \rightarrow \phi_{\ast,\text{true}} \ \ \ \text{or} \ \ \ \Phi_{\ast} \rightarrow \Phi_{\ast,\text{true}}
    \label{eq:phi_true}.
\end{equation}

Figure \ref{fig:mass_assembly_obs_vs_deconv} shows the resulting assembly history of progenitors at $z=0$ for galaxies in the mass range  $M_\ast = 10^{9}  - 10^{11.5}M_\odot$. The dotted lines represent evolutionary tracks based on the observed GSMF, while the solid lines correspond to the true GSMF, which accounts for random errors. As expected, the tracks derived from the true GSMF predict smaller stellar masses at a given redshift compared to those obtained from the observed GSMF. This effect is more pronounced at redshifts greater than $z\sim2$, where random errors can be as large as $\sim0.25$ dex. As an example, note that the track inferred for a progenitor of a galaxy with $M_{\ast}\sim3\times 10^{11} M_\odot$ at $z\sim0$ has a mass of $\sim2.5\times10^{9} M_\odot$ at $z\sim8$, while using the true GSMF, this mass is approximately $\sim1.6$ times smaller, or equivalently $\sim 0.19$ dex lower. 

Figure \ref{fig:numb_density_obs_vs_deconv} presents the number densities corresponding to the progenitors of the galaxies for which their assembly histories were shown in Fig. \ref{fig:mass_assembly_obs_vs_deconv}. In other words, this is the standard matching number density approach using the cumulative GSMF, $n_{\rm gal}$. We show the case for the observed GSMFs (dotted lines) and the true GSMFs (solid lines), which account for random errors. 
Specifically, if $M_{\ast,\text{true}}(z)$ is the track from the true GSMFs, we evaluate $n_{\text{gal,obs}}(M_{\ast,\text{true}})$. Note that the cumulative number densities inferred by $M_{\ast,\text{true}}(z)$ generally imply that galaxies originate from higher \emph{observed} cumulative number densities. 

Finally, although using $\phi_{\ast,\text{true}}$ represents a logical improvement to the number density matching technique, it does not account for the scatter in the mass assembly history of galaxies or the effects of mergers. To address this, we will employ a simple technique based on the assembly history of dark matter haloes and their evolution in number densities. 

\begin{figure}
		\includegraphics[width=\columnwidth]{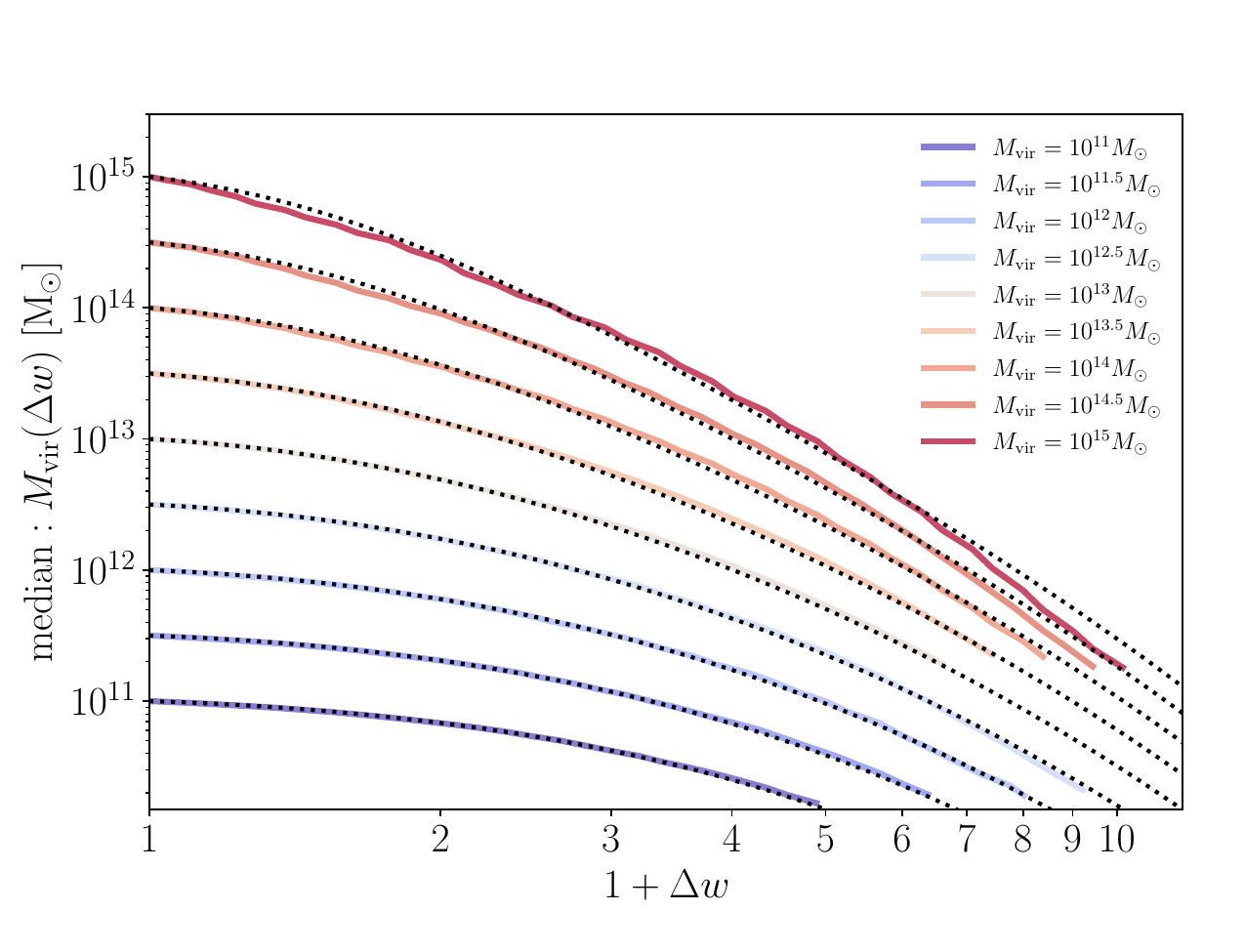}
			\caption{Median assembly history of dark matter haloes as a function of the time variable $\Delta \omega(z) = \omega(z) - \omega(z_0)$ with $\omega(z) = \delta_c / D(z)$. Median assembly histories for the BolshoiP simulation are shown for haloes in the range $M_\text{vir} = 10^{11}-10^{12.5}$ while the MDPL covers haloes in the range $M_\text{vir} = 10^{13}-10^{15}$. The dotted lines present the best-fit model to the simulations as described in the text. 
 		}
	\label{fig:median_halo_assembly}
\end{figure}

\begin{figure*}
		\includegraphics[width=\textwidth]{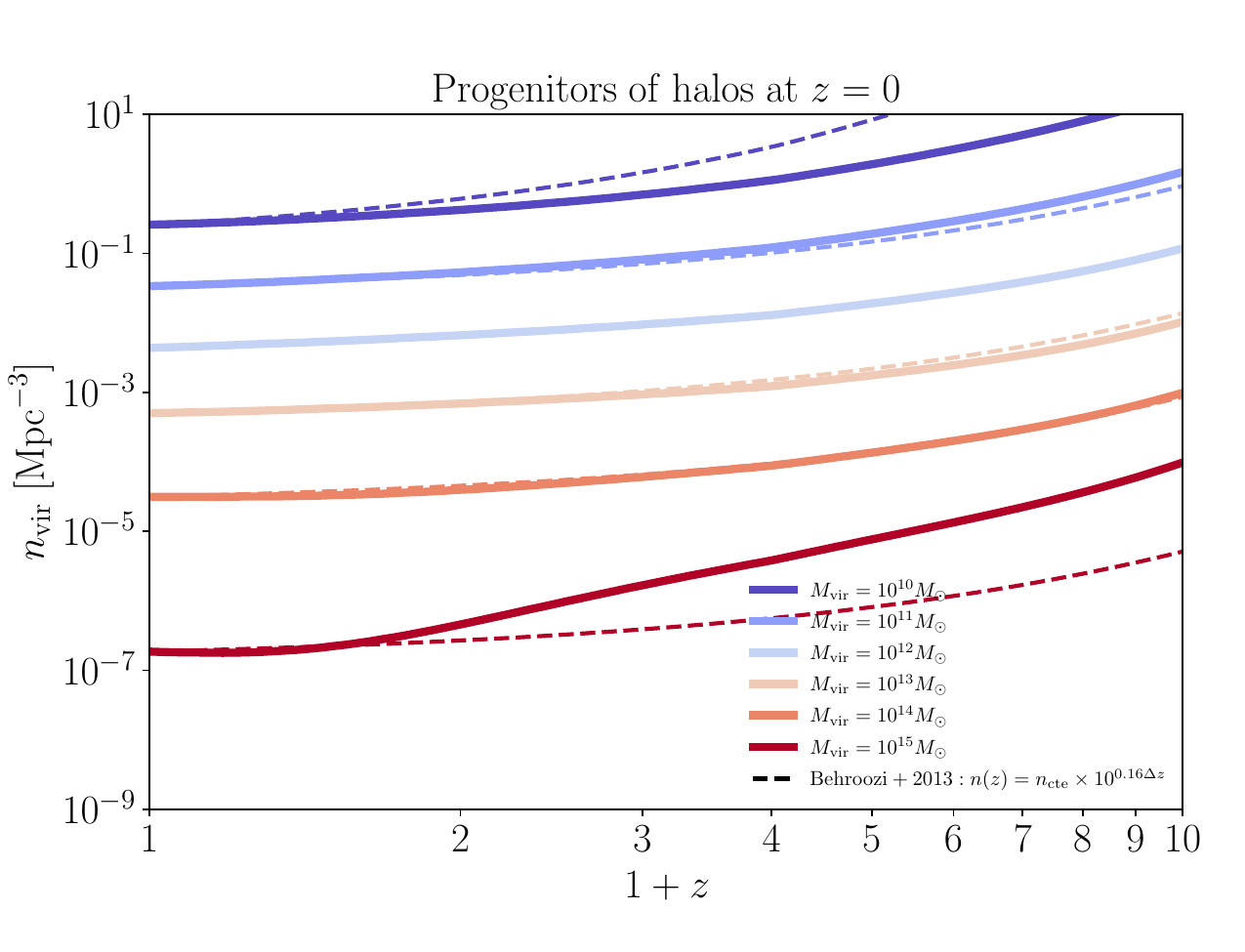}
			\caption{Evolution of cumulative number density for progenitors of dark matter haloes with final masses of $M_\text{vir} = 10^{10}, 10^{11}, 10^{12}, 10^{13}, 10^{14},$ and $10^{15} M_\odot$. The cumulative number densities are far from being constant, except for the most recent history ($z \lesssim 0.3$) of the $M_\text{vir} = 10^{15} M_\odot$ halo. The slopes for each curve are 0.22, 0.18, 0.15, 0.14, 0.17, and 0.33 dex per $\Delta z$ between $z=2$ and $z=4$, respectively.  The dashed lines presents the prescription of increasing the number density by $0.16 \Delta z$ dex by \citet{Behroozi+2013f}, which are consistent with our resulting cumulative number density tracks in the range of masses between $10^{11}$ and $10^{14} M_\odot$, while at the extrem ends, the disagreement becomes more evident for $z\gtrsim1$.
 		}
	\label{fig:halo_numb_density}
\end{figure*}

\begin{figure}
		\includegraphics[width=\columnwidth]{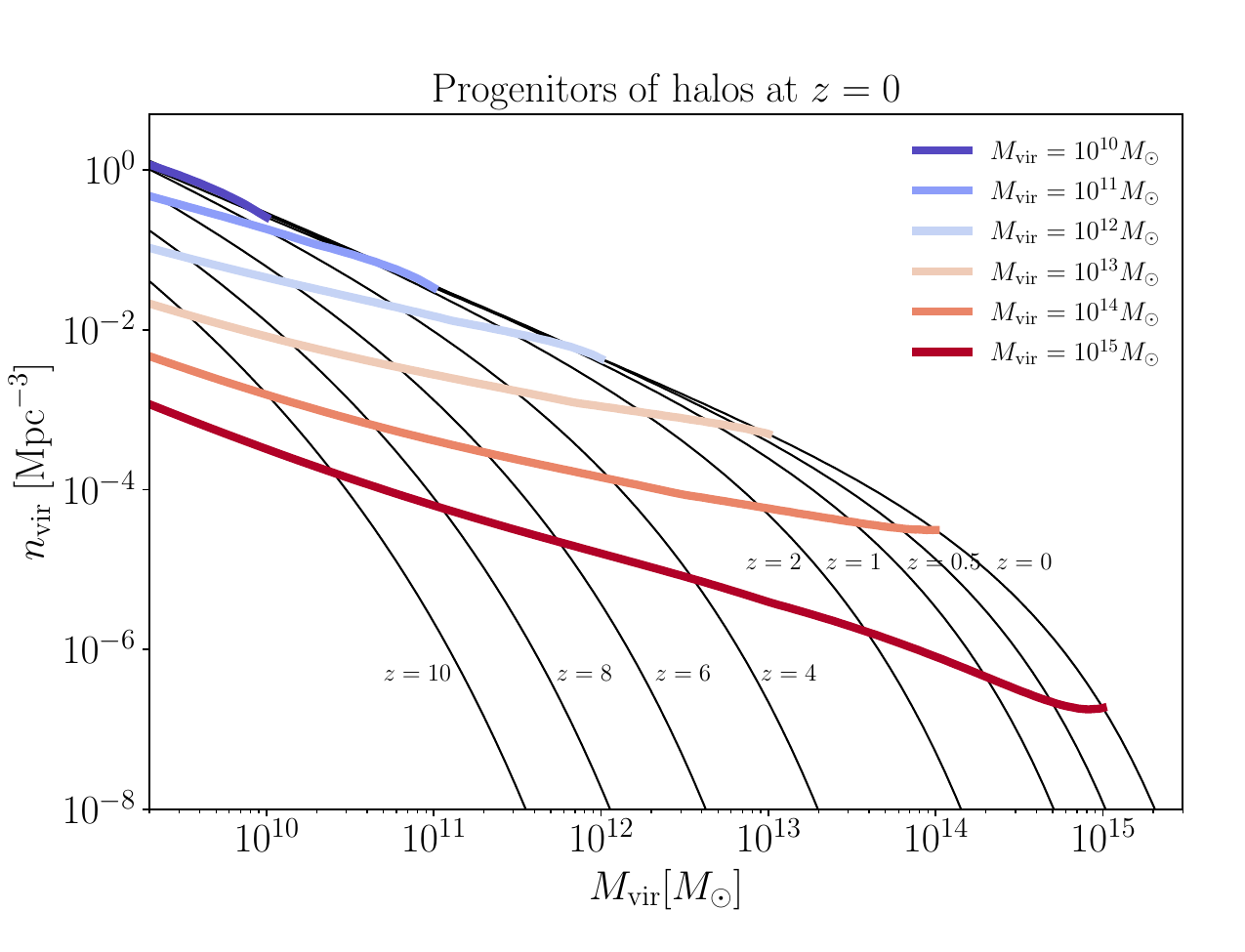}
			\caption{Median trajectories along the redshift evolution of the cumulative halo mass function for progenitors of dark matter haloes with final masses of $M_\text{vir} = 10^{10}, 10^{11}, 10^{12}, 10^{13}, 10^{14},$ and $10^{15} M_\odot$. We present the cumulative halo mass function at redshifts $z=0, 0.5, 1, 2, 4, 6, 8,$ and $10$. Similar to Figure \ref{fig:halo_numb_density}, this figure highlights that the number densities of halo progenitors are far from constant.  
 		}
	\label{fig:halo_mass_function}
\end{figure}

\begin{figure*}
		\includegraphics[width=\columnwidth]{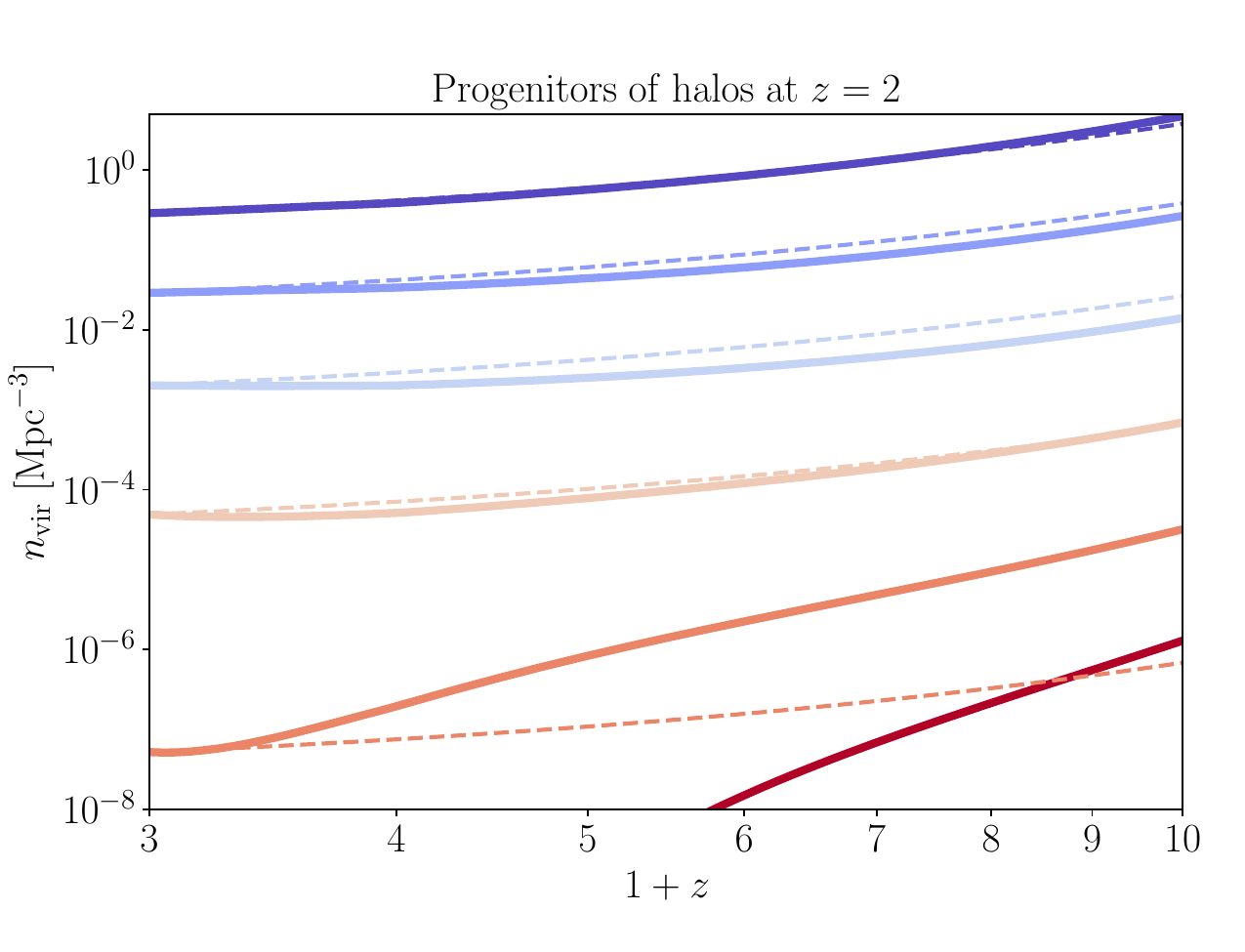}
		\includegraphics[width=\columnwidth]{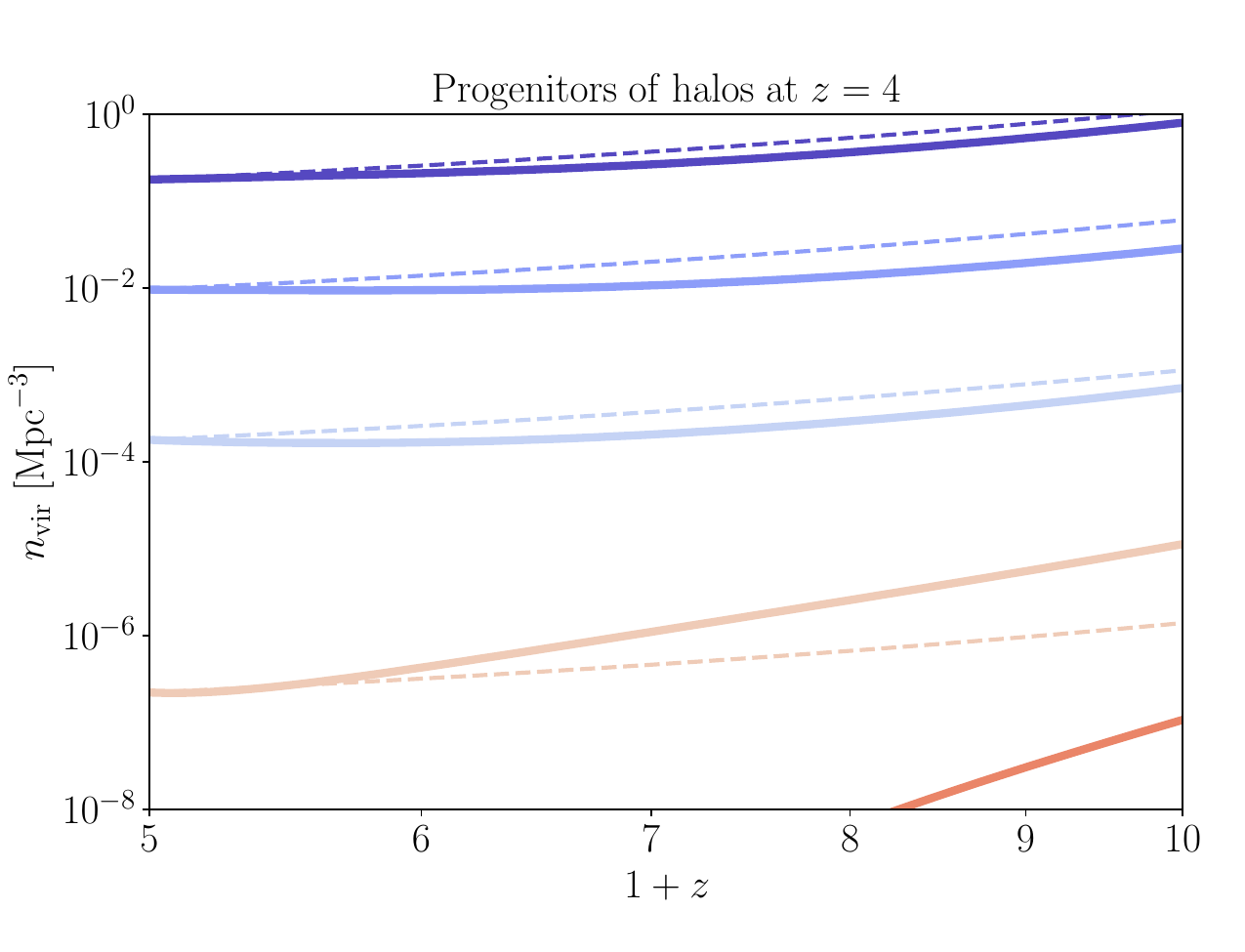}
			\caption{
			Same as Figure \ref{fig:halo_numb_density} but for progenitors at $z=2$ and $z=4$, shown in the left and right panels, respectively. Similar slopes are inferred for progenitors at $z=0$. However, the curves tend to flatten as the final mass of the progenitor is taken at higher redshifts and for lower-mass haloes. This indicates that the assumption of constant cumulative numbers is more accurate at higher redshifts. Similarly to Figure \ref{fig:halo_numb_density}, the dashed lines presents the prescription of increasing the number density by $0.16 \Delta z$ dex by \citep{Behroozi+2013f}. This prescription disagrees for most of halo cumulative number densities, except for $M_\text{vir}=10^{10}M_\odot$ and $M_\text{vir}=10^{13}M_\odot$ at $z=2$.   
 		}
	\label{fig:halo_numb_density_at_z2&z4}
\end{figure*}

\section{Evolving halo number densities inferred from Halo assembly histories}
\label{sec:evol_halo_nvir}

In the standard $\Lambda$CDM cosmological framework, galaxies form and evolve within extended dark matter haloes. The core idea we employ here is to use this framework, combined with the median assembly history of dark matter haloes, to trace back the most probable cumulative number densities of the progenitors of galaxies at a given redshift $z_0$, but see Section \ref{sec:generaliztion_of_the_approach} for a generalization to different types of galaxies. To understand how the number density of progenitor galaxies at $z_0$ evolves over cosmic time, it is reasonable to first examine this evolution in dark matter haloes.

\subsection{Halo assembly histories}
\label{sec:mah}

We use N-body simulations to derive the assembly histories of dark matter haloes. We use the N-body Bolshoi-Planck \citep[BolshoiP,][]{Rodriguez-Puebla+2016,Klypin+2016}, and the MultiDark Planck \citep[MDP,][]{Klypin+2016} cosmological simulations. All simulations are based on the $\Lambda$CDM cosmology with parameters consistent with the results from the \citet{Planck+2015}. The BolshoiP simulation has $2048^3$  particles of mass $1.9 \times 10^8 M_\odot h^{-1}$, in a box of side length $L_\text{ BP} = $ 250 $h^{-1}$Mpc, while the MDP has $3048^3$  particles of mass $1.5 \times 10^9 M_\odot h^{-1}$, in a box of side length $L_\text{ MDP} = $ 1 $h^{-1}$Gpc. Haloes/subhaloes and their merger trees have been characterized with the phase-space temporal halo finder \rockstar\ \citep{Behroozi+2013d} and their corresponding merger trees from \ctrees\ \citep{Behroozi+2013b}. Halo masses were defined using spherical overdensities according to the redshift-dependent virial overdensity $\Delta_\text{vir}(z)$ given by the spherical collapse model, with $\Delta_\text{vir} \sim 333$ at $z=0$ and $\Delta_\text{vir} = 178$ as $z\rightarrow \infty$. 

In this paper, we update the assembly histories inferred by \cite{Rodriguez-Puebla+2016} to make them as general as possible. To derive the assembly history of dark matter haloes, we use a time variable, $\omega$, that is invariant with respect to cosmology: the ratio $\delta_c(z)/D(z)$, where $\delta_c(z) = 1.686 \; \Omega(z)^{0.0055} \approx 1.686$ has a weak dependence on redshift and $D(z)$ is the linear growth rate. We define $\omega = \delta_c(z)/D(z)$, and thus $\Delta \omega = \omega(z) - \omega(z_0)$, with $z > z_0$. We assume that the general shape of the assembly history of dark matter haloes follows a formula similar to that constrained by \cite{Rodriguez-Puebla+2016} and proposed by \cite{Behroozi+2013}, but instead of using the redshift, $z$, as time variable we use $\Delta w$. We find that the median halo mass assembly is given by:
	\begin{equation}
		M_\text{ vir}(M_{\text{vir},0}, \Delta \omega) = 10^{13} m_\text{ 13} (\Delta \omega) 
        10 ^{f(M_{\text{vir},0}, \Delta\omega)} \; M_{\odot} h^{-1}_\text{ BP},
		\label{ec:median_mpeak_assembly}
	\end{equation}
where 
	\begin{equation}
		m_\text{ 13} (\Delta \omega) = (1+\Delta \omega)^{1.529}\left(1+\frac{\Delta \omega}{2}\right)^{-3.409}\exp{(-0.404 \Delta \omega)},
	\end{equation}
and the remaining terms are:	
	\begin{equation}
		f(M_\text{vir},\Delta \omega) = \log \left( \frac{M_{\text{vir},0}}{10^{13}M_{\odot} h^{-1}_\text{ BP}} \right)\frac{g(M_{\text{vir},0},1)}{g(M_{\text{vir},0},\Delta \omega)},
	\end{equation}
	\begin{equation}
		g(M_{\text{vir},0},\Delta \omega) = 1 + \exp\left[-4.08 \left(a_{\omega} -a_0(M_{\text{vir},0}) \right) \right],
	\end{equation}
and
	\begin{equation}
		a_0(M_{\text{vir},0}) = 0.286 - \log\left[ \left( \frac{10^{11.994}M_{\odot} h^{-1}_\text{ BP}}{M_{\text{vir},0}}\right)^{0.143} + 1 \right],
	\end{equation}
where we define a ``scale factor'' with $\Delta w$: $a_{\omega} = 1 / (1+\Delta\omega)$.

Figure \ref{fig:median_halo_assembly} shows the assembly history of dark matter haloes as a function of $\Delta \omega$ in the range of $M_\text{vir} = 10^{11} - 10^{15} M_{\odot}$. This figure presents the median assembly history from the BolshoiP simulation for haloes in the range $M_\text{vir} = 10^{11} - 10^{12.5} M_{\odot}$, while the MDPL simulation covers haloes in the range $M_\text{vir} = 10^{13} - 10^{15} M_{\odot}$. The dotted lines indicate the best fit to the simulations. While only selected results are plotted in Figure \ref{fig:median_halo_assembly}, we computed median assembly histories every 0.1 dex in halo mass, with a bin width of 0.25 dex. For BolshoiP, median assembly histories were derived for masses within the range $M_\text{vir}=10^{10.8} - 10^{14.9} M_\odot$, and for MDPL within $M_\text{vir}=10^{11.8} - 10^{15.2} M_\odot$. Observe the high degree of accuracy in our fits for most values of $\Delta \omega$. 

We are now prepared to analyze how the progenitors of dark matter haloes at $z=0$ evolved in their cumulative number density over time. Recall that our hypothesis is that galaxies form and evolve in dark matter haloes, and their evolutionary tracks should, on average, follow a similar pattern as their host haloes, as we will discuss below. 

\begin{figure*}
		\includegraphics[height=3.06in,width=7in]{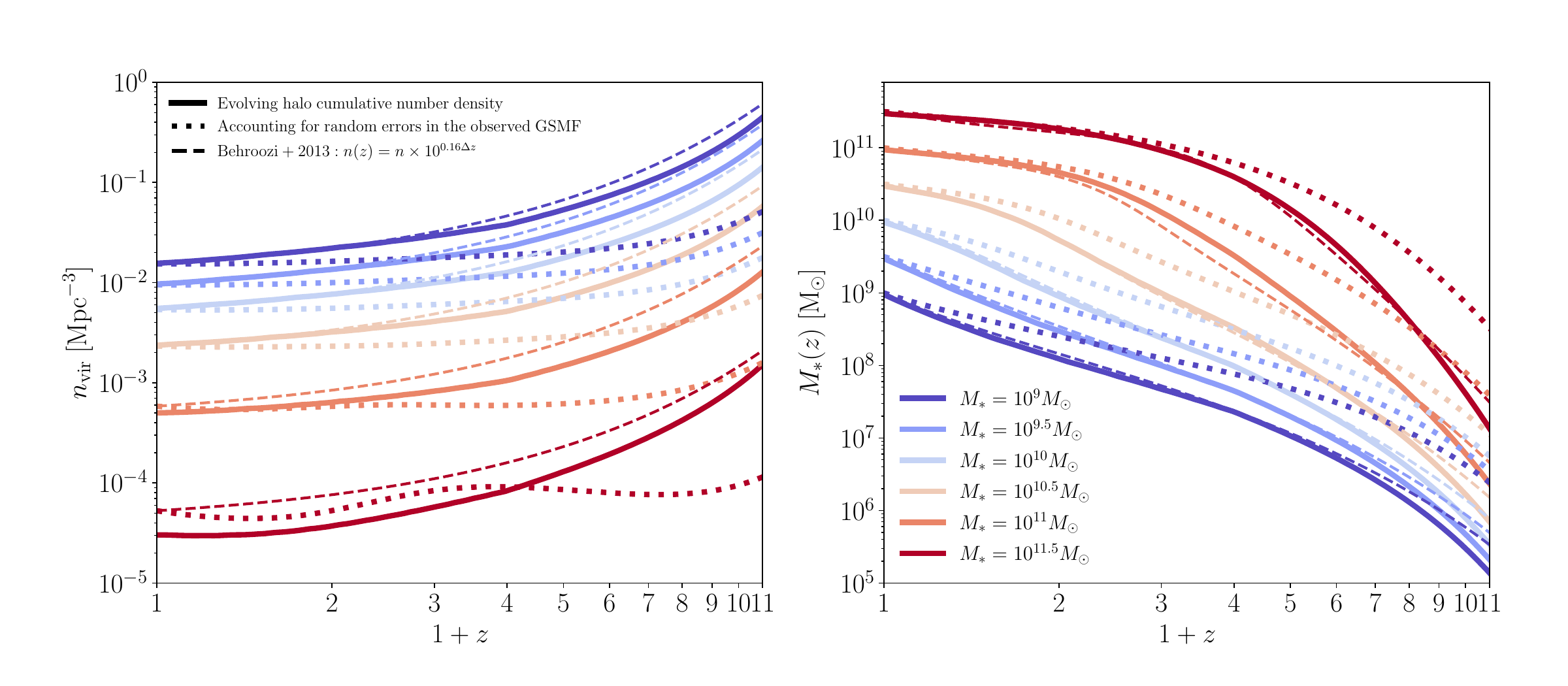}
			\caption{Differences between the cumulative number densities (left panel) associated with median galaxy tracks (right panel) inferred via the standard constant cumulative number density approach (but taking into account random errors; dotted lines in both panels) and those obtained by using evolving halo cumulative number densities and taking into account the intrinsic scatter around the stellar-to-halo mass relation (solid lines in both panels). The dashed lines indicate the resulting evolution from increasing the cumulative number density by 0.16$\Delta z$ dex, as suggested by \citep{Behroozi+2013f}. Note that this correction is applied to the observed GSMF, as described by the authors.
 		}
	\label{fig:galaxy_tracks_from_halo_num_tracks}
\end{figure*}

\subsection{Evolutionary Tracks of Halo Cumulative Number Densities}
\label{sec:evl_track_nhalo}

If we represent the assembly history of a dark matter halo by the quantity $M_\text{vir}(z|M_{\text{vir},0},z_0)$ where $M_{\text{vir},0}$ is the final halo mass at redshift $z_0$, then the associated evolution in their cumulative number density will be $n_\text{vir}(>M_\text{vir}(z|M_{\text{vir},0},z_0),z)$. In this way, we can trace the evolution of the cumulative number density of dark matter haloes if we know the median progenitor of these haloes, as shown below. To trace the evolution of the dark matter halo mass function, we use the updated version of the \hmf{} package \citep{Murray+2013}.\footnote{\url{https://hmf.readthedocs.io/}.} Specifically, we utilize the halo mass function implementation from \cite{Behroozi+2013} within \hmf{}, which represents an update to the \cite{Tinker+2008} model. We note that using other halo mass functions \citep[e.g.,][]{Rodriguez-Puebla+2016} will lead to similar conclusions.

Figure \ref{fig:halo_numb_density} illustrates how the median progenitors of dark matter haloes at $z=0$ from Figure \ref{fig:median_halo_assembly} evolve in their cumulative number density over time. As shown in the figure, cumulative number densities are far from constant, except for the most recent history of progenitors of haloes with $M_\text{vir} = 10^{15}M_\odot$, where the number density becomes nearly constant after redshift $z\lesssim 0.3$. The behavior of the curves in Figure \ref{fig:halo_numb_density} suggests that the diversity in halo assembly history and mergers leads to significant changes in their number densities. To quantify these changes, we computed the slope within the range of  $z=1$ and $z=3$ where most of the curves show an increase, and found slopes of $0.22, 0.18, 0.15, 0.14, 0.17$ and $0.33 \Delta z$ dex, respectively for progenitors at $z=0$ with masses $M_\text{vir} = 10^{11}, 10^{12}, 10^{13}, 10^{14}$ and $10^{15} \; M_{\odot}$. These values are consistent with the average change in number density of 0.16 dex per $\Delta z$ dex found by \cite{Behroozi+2013f}. To verify this, the dashed lines in Figure \ref{fig:halo_numb_density} show the change in number density using the \citet{Behroozi+2013f} prescription. Notice the excellent agreement with haloes in the range $M_\text{vir}\sim10^{11}-10^{14}M_\odot$, while at the extreme ends, the disagreement becomes more evident for $z\gtrsim1$. While these authors quantified the change in number density using SHAM in the same $N$-body simulations as ours, this exercise of showing the number density evolution of the haloes first demonstrates that they were \emph{effectively} measuring the change in the number density of dark matter haloes. 

Similar to Figure \ref{fig:halo_numb_density}, Figure \ref{fig:halo_mass_function} illustrates the redshift evolution of the halo mass function and how dark matter haloes evolve over time. This figure emphasizes that progenitors of $z=0$ haloes originated from significantly higher number densities at earlier redshifts. The most dramatic case is for progenitors of $M_\text{vir} = 10^{15} M_{\odot}$ haloes, which have experienced a decrease of approximately four orders of magnitude in their number densities. The behaviors observed for haloes are expected to be inherited by their associated galaxies through the galaxy-halo connection described below in Section \ref{sec:proj_halo_den_onto_gal_den}. 

Figure \ref{fig:halo_numb_density_at_z2&z4} presents similar information to Figure \ref{fig:halo_numb_density}, but for progenitors of haloes at redshifts $z=2$ and $z=4$. While we observe similar slopes to those of progenitors at $z=0$, we also notice that the curves tend to flatten as they approach the redshift of the progenitor. This suggests that at higher redshifts, the assumption of constant cumulative number density holds over a wider range and becomes more accurate closer to the redshift of the progenitors. The dashed lines again show the approximation from \citet{Behroozi+2013f}. Notice that there is disagreement for most halo trajectories except those with $M_\text{vir}=10^{10}M_\odot$ and $M_\text{vir}=10^{13}M_\odot$ at $z=2$, suggesting that the \citet{Behroozi+2013f} approximation is less valid at higher redshifts. 

We conclude this section by stating that while a model of the form $\log n(z)\sim \alpha \Delta z$, where $\alpha$ is a constant slope with mass, captures the trend of evolving cumulative number densities, mostly for intermediate masses and for progenitors defined at low $z$, using the full calculation will result in more accurate evolutionary tracks. Additionally, the slope $\alpha$ likely depends on cosmological parameters, as both the halo mass function and halo merger trees are influenced by them.

Next, we explore how the evolutionary tracks presented in Figures \ref{fig:halo_numb_density} and \ref{fig:halo_numb_density_at_z2&z4} are projected into galaxy evolutionary tracks. 

\section{Semi-Empirical Approach: Projecting evolving halo number densities onto galaxy number densities}
\label{sec:proj_halo_den_onto_gal_den}

In the previous section, we explored how the progenitors of dark matter haloes at a given observed redshift, $z_0$, evolve in their cumulative number densities using median halo assembly histories. The central idea in this section is to establish the galaxy-halo connection by assuming a monotonic increasing stellar-to-halo mass relation with some associated scatter. We will derive this relation using the traditional SHAM method \citep{Frenk+1988,ValeOstriker2004,Wechsler+2018}.\footnote{\citet{Frenk+1988} are perhaps the earliest authors to introduce SHAM in its current form and to empirically study the galaxy-halo connection; see their Figure 12.} This approach enables us to project evolving halo cumulative number densities into evolving galaxy number densities and, ultimately, galaxy stellar masses. $M_\ast$.

\subsection{Galaxy-halo connection}

We begin by proposing that the true GSMF results from convolving an intrinsic GSMF with a Gaussian distribution, $\mathcal{H}$, where the width of the Gaussian represents the scatter around the stellar-to-halo mass relation \citep{Behroozi+2013,Rodriguez-Puebla+2017}. This approach is analogous to Eq. (\ref{eq:gsmf}), but takes the following form:
\begin{equation}
    \phi_{\ast,\text{true}}(M_\ast) = \int_{-\infty}^{\infty} \mathcal{H}(x-\log M_{\ast}) \, \phi_{\ast,\text{intr}}(x) d x.
    \label{eq:intrinsci_GSMF}
\end{equation}
Here we assume that the scatter around the stellar-to-halo mass relation is of $0.15$ dex at all redshifts consistent with previous semi-empirical works \citep[see e.g.,][see also \citealp{Porras-Valverde+2023}]{Yang+2012,Rodriguez-Puebla+2015,Wechsler+2018,Behroozi+2019}. Although evidence suggests that scatter may increase at lower masses (see references above), we do not model this effect here. Nevertheless, larger scatter at lower masses has a minor impact, and including it would only reinforce our conclusions. The next step is to establish the relationship between $\phi_{\ast,\text{intr}}$ and the halo mass function, $\phi_{\text{vir}}$, using the traditional SHAM technique \citep[SHAM][]{ValeOstriker2004,Conroy+2009}:
\begin{align}
        \int_{M_\ast}^{\infty}\phi_{\ast,\text{intr}}(x) \, dx = & 
        \int_{M_\text{DM}}^{\infty} \phi_{\text{DM}}(h) \, dh 
          \; \text{ or}  \,  \nonumber \\ 
        &  n_{\ast,\text{intr}} (>M_\ast) = n_\text{DM} (>M_\text{DM}), 
    \label{eq:SHAM}
\end{align}
where the subscript ``DM'' refers to the dark matter halo plus subhalo mass function and $M_\text{DM}$ is given by
\begin{equation}
	   M_\text{DM} = \left\{ 
			\begin{array}{c l}
				M_\text{vir} & \mbox{Distinct haloes}\\
				M_\text{peak} & \mbox{Subhaloes}
			\end{array},\right.
\end{equation}
and $M_\text{peak}$ is the maximum halo mass throughout the entire 
history of a subhalo. Similar combinations have been shown to reproduce the observed clustering of galaxies \citep{Reddick+2013,Calette+2021b,Kakos+2024} and the environment of central and satellite galaxies \citep{Dragomir+2018}, see \citet{Rodriguez-Puebla2024} for a discussion. Equation (\ref{eq:SHAM}) establishes the stellar-to-halo mass relation \citep[see e.g.,][]{Behroozi+2010}. We emphasise that using \emph{intrinsic} stellar masses is of extreme importance since those are the ones that \emph{should be compared} against dark matter haloes and will ultimately constrain the physics behind galaxy formation models. 

We again use the \cite{Behroozi+2013} halo mass function, $n_\text{vir}$, and correct for the small effect of subhaloes $\sim 20-25\%$ for the total halo mass function. Here we use a heuristic correction based on the subhalo mass function from \citet{Rodriguez-Puebla+2016}. That is, we parametrize the subhalo mass function in terms of the halo mass function given by
\begin{equation}
    n_\text{sub}(> M) = \mathcal{S}(M) \times n_\text{vir}(> M)
\end{equation}
such that
\begin{equation}
    n_\text{DM}(> M_\text{DM}) =  
    n_\text{vir}(> M_\text{DM})\left(1+\mathcal{S}(M_\text{DM})\right)
\end{equation}
To characterise $\mathcal{S}$ we use the best fit to the normalization for the subhalo mass function and the cut-off mass from \citet{Rodriguez-Puebla+2016} in such a way that the total subhalo mass is given by:
\begin{equation}
    \mathcal{S}(M_\text{peak})  = 
    1.78 \; \left( \frac{C_\text{sub}(z)}{C_\text{sub}(0)} \right) \; \exp \left[ -\left(\frac{ M_\text{peak}}{M_\text{cutoff}}\right)^{0.221} \right],
\end{equation}
where 
\begin{equation}
    \log \left( \frac{C_\text{sub}(z)}{C_\text{sub}(0)} \right) = 0.0087 z - 0.0113 z^2 - 0.0039 z^3 + 0.0004 z^4,
\end{equation}
and
\begin{equation}
    \log \left( \frac{M_\text{peak}}{M_\odot} \right)  = 11.905 - 0.636z - 0.020z^2+0.022z^3-0.001z^4.
\end{equation}

\subsection{Galaxy evolutionary tracks}
\label{sec:gal_evol_tracks}

To project evolutionary tracks from evolving halo number densities onto galaxies, we first compute $n_\text{DM}(>M_\text{vir}(z|M_{\text{vir},0},z_0),z)$ for a series of redshift snapshots (see above Section \ref{sec:evl_track_nhalo}). For simplicity, we will refer to this as $n_\text{DM}(>M_\text{vir},z|z_0)$. We then solve Eq. (\ref{eq:SHAM}) for $M_\ast$. In other words, we derive galaxy stellar mass evolutionary tracks by solving the following equation:
\begin{equation}
  n_{\ast,\text{intr}}(>M_\ast,z|z_0)  = n_\text{DM}(>M_\text{vir},z|z_0),
  \label{eq:gal_num_bs_halo_tracks}
\end{equation}
redshift by redshift to obtain the galaxy's evolutionary track, $M_\ast(z|M_{\ast,0},z_0)$.\footnote{ While previous authors have studied the impact of satellite galaxies on Eq. \ref{eq:gal_num_bs_halo_tracks} \citep[e.g.,][]{Neistein+2011,Rodriguez-Puebla+2012,Rodriguez-Puebla+2013,Niemiec+2019,Danieli+2023}, showing that corrections have to be made, here we will make the assumption, for simplicity, that satellites evolve similar to central galaxies, \citep{Simha+2012,Behroozi+2013,Rodriguez-Puebla+2013,Rodriguez-Puebla+2017}.}

The left panel of Figure \ref{fig:galaxy_tracks_from_halo_num_tracks} illustrates the differences between the galaxy tracks inferred from the empirical approach by matching cumulative number densities from the true GSMF (dotted lines, Eqs. \ref{eq:num_den} and \ref{eq:phi_true}, see also Figure \ref{fig:mass_assembly_obs_vs_deconv}) evaluated in the observed GSMF, and those obtained using the galaxy-halo connection by evolving halo cumulative number densities (solid lines) based on Eq. (\ref{eq:gal_num_bs_halo_tracks}). This comparison is made for galaxies with masses of $M_\ast = 10^{9}, 10^{9.5}, 10^{10}, 10^{10.5}, 10^{11}$, and $10^{11.5} M_\odot$ at $z=0$. The first notable observation is that at $z=0$, the cumulative number density is larger for the true GSMF at higher masses compared to the one based on halo number densities. This discrepancy arises because the true GSMF is the result of the convolution between the intrinsic GSMF and the intrinsic scatter around the stellar-to-halo mass relation, $\phi_{\ast,\text{true}} = \mathcal{H} \; \circ  \; \phi_{\ast,\text{intr}}$, see Eq. (\ref{eq:intrinsci_GSMF}). Therefore, the differences observed at $z\sim0$ are simply another manifestation of the Eddington bias, see Section \ref{sec:random_errors} and Eq. (\ref{eq:convGSMF_approx}) for a crude approximation. At higher redshifts, the differences are much larger and have a different origin. While the tracks based on the galaxy-halo connection, that is, the semi-empirical approach (solid lines), reveal the intrinsic evolution in the number density of haloes and their associated galaxies, the tracks inferred by matching galaxy cumulative number densities, i.e. the purely empirical approach (dotted lines), primarily reflect the difference between the true and observed GSMFs, as explained in Figure \ref{fig:numb_density_obs_vs_deconv}. 

The slopes observed for halo number densities will, by construction, be similar to those from dark matter haloes, as reported in Section \ref{sec:evl_track_nhalo}, which on average are of the order: 
\begin{equation}
    \left\langle \frac{d\log n_{\ast}}{dz} \right\rangle = 0.198 \text{ [dex]}.
\end{equation}
This is consistent with the proposed value of 0.16 dex by \citet{Behroozi+2013f}, shown as dashed lines in Figure \ref{fig:galaxy_tracks_from_halo_num_tracks}. However, we emphasize that these slopes depend on the final stellar mass at $z=0$; steeper slopes are expected in more massive galaxies. The discrepancy in normalization between our tracks and those inferred based on the \citet{Behroozi+2013f} approximation arises from the fact that these authors used the observed GSMFs instead of the intrinsic ones, evidencing again the effect of the \citet{Eddington1913,Eddington1940} bias. Consequently, their cumulative number densities are higher than ours for a given stellar mass. This also implies that the tracks inferred from our method and those based on \citet{Behroozi+2013f} will lead to the selection of different dark matter haloes.

The fact that all the curves have non-zero slopes suggests that scatter in mass assembly history and mergers are continuously affecting all galaxies. That is, \textit{galaxies are constantly changing their number density as they evolve across the time}. Specifically, at lower masses, $M_\ast\lesssim 3\times 10^{10} M_{\odot}$, mergers are expected to play a minor role in the evolution of these galaxies (less than $\sim5\%$, see Figure 13 from \citealp{Rodriguez-Puebla+2017}), with the main effect being the result of assembly histories. In contrast, at higher masses, both effects may contribute, with mergers perhaps playing the leading role since most massive galaxies are quiescent \citep{Rodriguez-Gomez+2016}.

The right panel of Figure \ref{fig:galaxy_tracks_from_halo_num_tracks} presents the evolutionary tracks of the galaxies discussed in the left panel. Generally, we observe that at higher redshifts, the intrinsic stellar masses of the galaxies (solid lines) are smaller compared to those based on matching galaxy cumulative number densities of the true GSMF (dotted lines) at a given redshift. 
For instance, at  $z\sim2$, the differences are a factor of $\sim3$ for galaxies with final stellar masses in the range $M_\ast = 10^{9}-10^{10.5} M_\odot$, while they are around 1.8 and 1.2 for galaxies with $M_\ast = 10^{11} M_\odot$ and $M_\ast = 10^{11.5} M_\odot$, respectively. At even higher redshifts, the differences become more pronounced; for example, at $z\sim7$, the differences can be around one order of magnitude or more across all mass ranges. In contrast, although our evolving cumulative number densities in the left panel differ from those obtained using the \citet{Behroozi+2013f} approach, the resulting galaxy evolution tracks are highly consistent with ours. This suggests that despite selecting different haloes, and therefore different number densities, the constant slope approach effectively produces reasonably galaxy tracks. 

As we will see below, however, the differences in the approaches will have significant implications for our understanding of how galaxies formed and evolved within dark matter haloes.

\begin{figure*}
		\includegraphics[width=\textwidth]{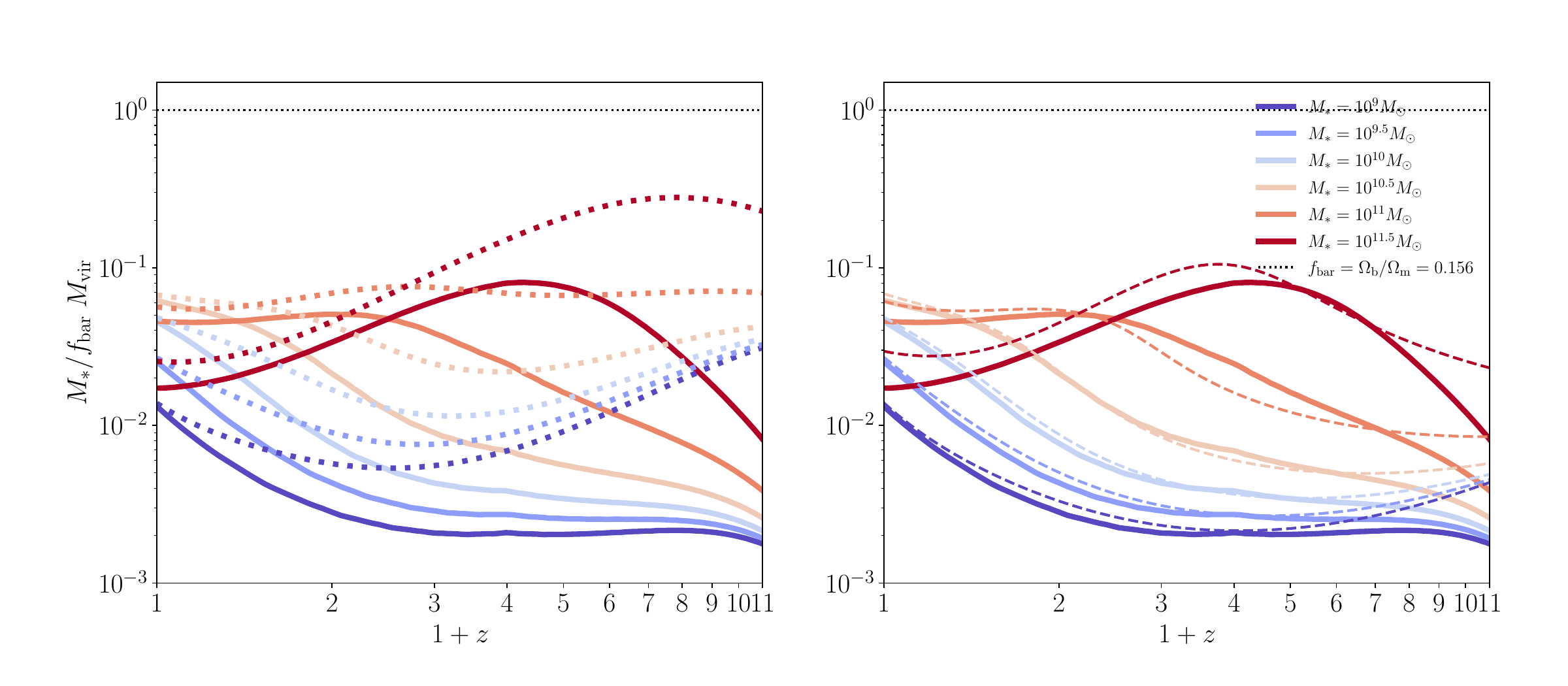}
			\caption{Median evolutionary tracks for the $M_\ast/f_{\rm bar}M_\text{vir}$ ratios of progenitors of haloes at $z=0$. The dotted lines in the left panel represent the ratios inferred by combining tracks from matching cumulative number densities between adjacent redshift of the observed GSMF and the median halo evolutionary tracks (see the text for details), while the solid lines (in both panels) correspond to the evolving halo cumulative number densities (Eq. \ref{eq:gal_num_bs_halo_tracks}). The thin dotted horizontal line  corresponds to $M_\ast/M_\text{vir}=f_{\rm bar}$, where the universal baryon fraction $f_\text{bar} = 0.156$. It is observed that the matching cumulative number density approach predicts higher stellar fractions in the past compared to the evolving halo cumulative number densities approach. This difference arises from the slower mass change in the progenitors inferred using the matching cumulative number density method. The right panel is similar to left panel but shows the tracks with the approximation of \citet[][dashed lines]{Behroozi+2013f}. 
        }
	\label{fig:imposible_galaxies}
\end{figure*}

\section{Discussion}
\label{sec:discussion}

In this section, in the context of the galaxy-halo connection, we begin by discussing the implications of different methodologies for inferring galaxy evolutionary tracks, namely: the constant number density approach, the increase of the number density of 0.16 de per $\Delta z$ proposed by \citet{Behroozi+2013f}, and the full approach presented here, which evolves the halo number density using median halo merger trees, it also takes into account random errors in stellar mass as well as the scatter in the stellar-to-halo mass relation. Additionally, we explore how our approach can be extended to consider the full diversity of halo assembly histories instead of just the median, and how it connects to other galaxy properties. We conclude this section by discussing an immediate application: using this framework to investigate the problem of impossible galaxies, i.e., massive galaxies that appear too large to be explained within the $\Lambda$CDM model.

\subsection{Implications for the galaxy-halo connection}

Figure \ref{fig:imposible_galaxies} shows the median evolutionary tracks for the $M_\ast /f_\text{bar} M_\text{vir}$ ratios of progenitors of the haloes at $z=0$ presented in Figure \ref{fig:galaxy_tracks_from_halo_num_tracks} for which their host galaxies correspond to a mass range between $M_\ast = 10^{9 }M_{\odot}$ and $10^{11.5} M_{\odot}$. The term $f_\text{bar}$ is the universal baryonic mass fraction; for the cosmological parameters used in this paper, $f_\text{bar} \equiv \Omega_\text{b}/\Omega_\text{m} = 0.156$. In the extreme case where the halo captures and retains all available baryons, and they all transform into stars, $M_\ast /f_\text{bar} M_\text{vir}=1$ (dotted line), that is, $M_\ast /M_\text{vir}=f_\text{bar}$, the baryonic mass fraction would reach its logical maximum. In the left panel of this figure, the solid lines result from solving Eq. (\ref{eq:gal_num_bs_halo_tracks}). In contrast, the dotted lines show the results of combining galaxy tracks by matching the cumulative number densities of the observed GSMF, see Section \ref{sec:num_match}, with the halo tracks described in Section \ref{sec:mah}. In this case, we relate galaxy tracks to halo tracks by matching their cumulative number densities only at $z=0$.

\begin{figure*}
		\includegraphics[width=\textwidth]{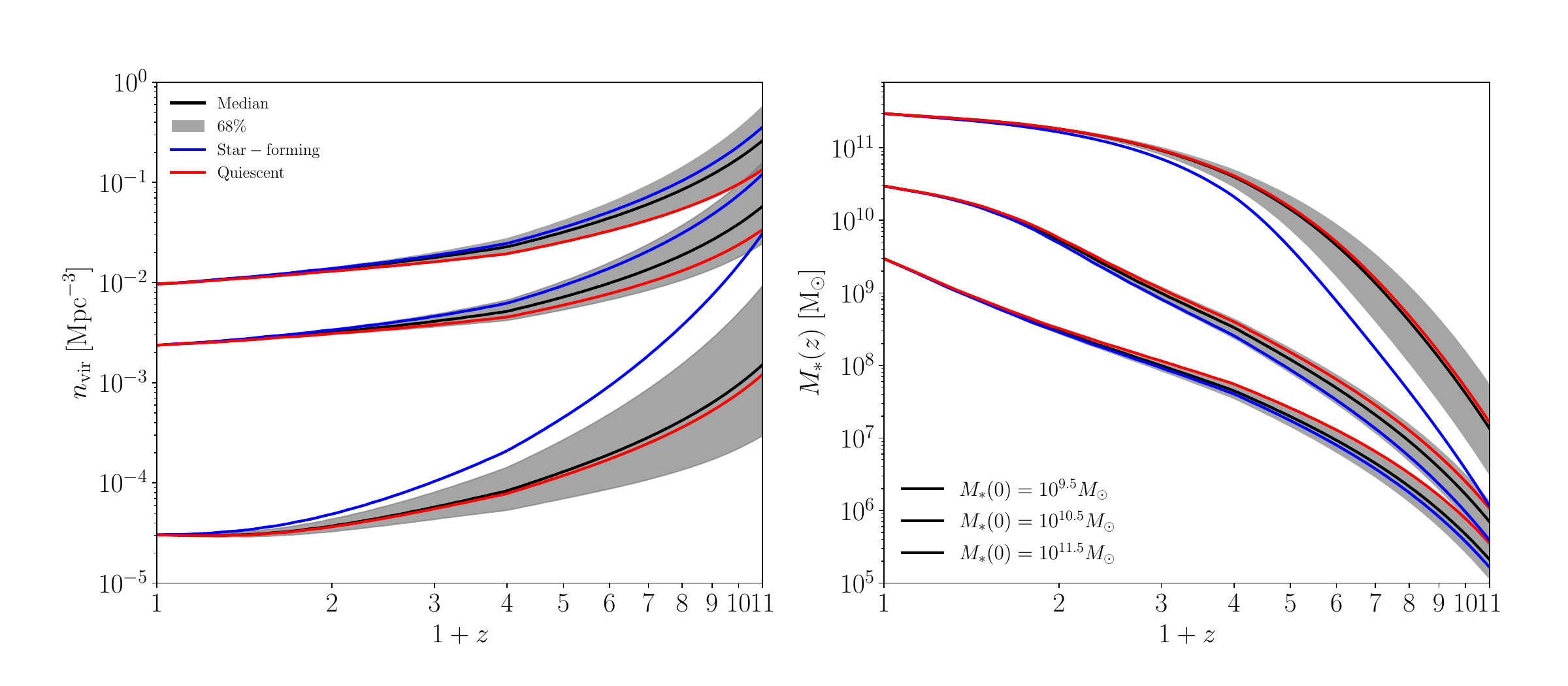}
			\caption{Evolving halo cumulative number densities (left panel) for the progenitors of galaxies with masses $M_\ast = 10^{9.5},10^{10.5}$ and $10^{11.5}M_\odot$  and their corresponding assembly histories (right panel). The solid lines represent the medians, while the shaded areas indicate the $68\%$confidence interval of the progenitor distribution. The blue and red lines correspond to star-forming and quiescent galaxies, respectively, based on the quiescent fraction from \citet{Moustakas+2013}. Star-forming galaxies exhibit median assembly histories that differ significantly from those of quiescent galaxies, as demonstrated by this implementation of our extended model.}
	\label{fig:generalization_to_the_method}
\end{figure*}

We observe that the median $M_\ast/f_\text{bar}M_\text{vir}$ ratios for both derivations are much lower than 1 at all redshifts.  However, for the tracks inferred from the observed GSMF, at $z\sim7-8$ the progenitors of $M_\ast = 10^{11.5}M_\odot$ galaxies at $z=0$ have ratios that are about 0.3. In other words, these are close to the extreme case that all baryons are captured and transformed into stars. In contrast, the progenitors of the same galaxies, as inferred from Eq. (\ref{eq:gal_num_bs_halo_tracks}), have much smaller ratios at the same redshifts. As seen in Figure \ref{fig:imposible_galaxies}, the maximum stellar-to-baryonic halo mass median ratios attained by massive galaxies is $\sim 0.1$. In general, the ratios for all masses are always smaller when using the evolving cumulative number density of haloes instead of a constant one and without taking into account random errors as in the purely empirical approach.

The above leads us to conclude that while the observed GSMFs suggest galaxies with higher baryon efficiencies, when accounting for random errors and the scatter in the stellar-to-halo mass relation, this tension is significantly reduced. We note, however, that these results refer to median galaxy tracks. At redshifts higher than $z\sim3$, random errors are $\sim0.4$ dex, and if we assume an intrinsic scatter in the stellar-to-halo mass relation of about $0.15$ dex, then the observed galaxies would be scattered up and down by $\sigma_\text{tot}\sim 0.43$ dex, assuming both deviations follow a Gaussian distribution. Interestingly, at the redshift ranges $z\sim3-8$, galaxies that are $\sim 3-4 \sigma_\text{tot}$ above the median will have observed stellar masses close to or even \emph{exceeding} the universal baryon fraction. Conversely, at the same redshift, some galaxies could also have very low $M_\ast/M_\text{vir}$, as discussed in more detail below in Section  \ref{sec:jwst_massive_gals}.

Finally, the right panel of Figure \ref{fig:imposible_galaxies} compares our evolving cumulative number density approach to that of \citet{Behroozi+2013f}. The differences between the two methods are better understood as a function of galaxy mass. At lower masses, their approach closely resembles ours, which is a direct consequence of the results shown in Figure \ref{fig:halo_numb_density}. For low-mass haloes (our smallest galaxy with $M_\ast=10^{9}M_\odot$ corresponds to a halo of $M_\text{vir}\sim10^{11}M_\odot$), the \citet{Behroozi+2013f} approach aligns well with our more general evolving cumulative number density method. However, at higher masses, a larger discrepancy emerges. This discrepancy is not just due to mass differences. The \citet{Behroozi+2013f} approach was designed to work with the \emph{observed} GSMF, whereas our method applies to the \emph{intrinsic} GSMF. As a result, part of the difference arises because, in the \citet{Behroozi+2013f} approach, Eddington bias plays a significant role. Consequently, applying the \citet{Behroozi+2013f} approach may lead to inconsistencies with the SHMR.

\subsection{Generalization to include different types of galaxies}
\label{sec:generaliztion_of_the_approach}

Galaxies exhibit a bimodal distribution in key properties such as color, star formation rate, and morphology. A natural question arises: how can we generalize our method to account for different galaxy types, given that much of our discussion has focused on median evolutionary tracks for all haloes/galaxies? For the following discussion, we implicitly assume that the SHMR depends only on halo mass. However, we acknowledge that this assumption may not be entirely realistic, as previous empirical studies have shown that the SHMR varies with galaxy color \citep{Mandelbaum+2006, More+2011, Rodriguez-Puebla+2015,Mandelbaum+2016}, star formation activity \citep{Kakos+2024}, and morphology \citep{Correa_Schaye2020}.

To generalize our discussion, we define the probability distribution function $\mathcal{P}(M_\text{vir},z|M_{\text{vir},0},z_0)$ as the conditional probability that a halo with a final mass $M_{\text{vir},0}$ at $z_0$ has a progenitor of mass $M_\text{vir}$ at $z$ with a median distribution characterized by Eq. (\ref{ec:median_mpeak_assembly}). While several studies have attempted to constrain this distribution using different approaches \citep[e.g.,][]{Wechsler+2002, Yang+2011, vandenBosch+2014}, here we adopt a simplified model to illustrate its implementation as discussed below.

The model by \citet{Wechsler+2002}, found that halo mass growth can be parametrizes as by the following functional form:
\begin{equation}
    M_\text{vir}(a|M_\text{vir,0},a_\text{form}) = M_\text{vir,0}\, \exp \left[ -\gamma a_\text{form}\left( \frac{a_0}{a} -1 \right) \right],
    \label{eq:assembly_histo_w02}
\end{equation}
where $a$ is the scale factor, and $a_\text{form}$ is the formation redshift of haloes, given by $a_\text{form} = \beta/c_\text{vir}$ with $c_\text{vir}$representing the dark matter halo concentration. While Eq. (\ref{ec:median_mpeak_assembly}) provides a more accurate description of the median evolution of dark matter haloes \citep[see][for caveats regarding exponential fits]{Behroozi+2013}, the equation above helps generalize our formalism by characterizing the distribution of formation times in dark matter haloes. In particular, since halo mass is fixed, this equation provides a distribution of halo mass assembly histories with different formation times. We define the following ratio:
\begin{equation}
    \mathcal{A} \equiv \frac{M_\text{vir}(a|M_\text{vir,0},a_\text{form})}{M_\text{vir}(a|M_\text{vir,0},\bar{a}_\text{form})}
    = \exp \left[ -\gamma(\Delta a_\text{form})\left( \frac{a_0}{a} -1 \right) \right],
    \label{eq:assembly_ratio}
\end{equation}
where $\Delta a_\text{form} = a_\text{form} - \bar{a}_\text{form}$. This equation quantifies differences in assembly histories between the median track, characterized by the formation time $\bar{a}_\text{form}$ and haloes with different formation times  $a_\text{form}$. We can thus generalize Eq. (\ref{ec:median_mpeak_assembly}) for our assembly history as follows:
\begin{equation}
    M_\text{vir}(M_\text{vir,0},\Delta w,a_\text{form}) = 
    M_\text{vir}(M_\text{vir,0},\Delta w) \mathcal{A}(\Delta a_\text{form},a).
\end{equation}
Additionally, notice that in Eq. (\ref{eq:assembly_ratio}), as $a\rightarrow a_0$ then $\mathcal{A}\rightarrow 1$. In Rodriguez-Puebla et al. (in prep) we will show that the distribution of halo assembly histories are well represented by the values of $\gamma=1$ and $\beta=2.26$.   

Notice that the dependence of formation times in Eq.  (\ref{eq:assembly_ratio}) on halo concentration provides a way to characterize $\mathcal{P}(M_\text{vir},z|M_{\text{vir},0},z_0)$ from the conditional distribution $\mathcal{P}(c_\text{vir}|M_{\text{vir},0})$; which is approximately lognormal with a width of $\sim0.1$ dex \citep{Neto+2007,Dutton+2014,Diemer_Kravtsov2015}. Specifically, in this paper, we adopt the $c_\text{vir}-M_\text{vir}$ relation from \citet{Dutton+2014} assuming a lognormal distribution with an intrinsic scatter of $\sigma_\text{c}=0.11$ dex. Once the concentration distribution is defined, we can trace the evolutionary track of any dark matter halo based on its concentration.

Next, we assume that a dark matter halo’s assembly history governs the assembly history of its galaxy \citep{Dekel+2014,Mitra+2015,Rodriguez-Puebla+2016a}, correlating the galaxy’s star formation activity with the halo concentration, similar to the age-matching model by \citet{HearinWatson2013}.

Figure \ref{fig:generalization_to_the_method} reproduces the median evolving halo cumulative number density for the progenitors of galaxies with masses $M_\ast=10^{9.5}, 10^{10.5}$ and $10^{11.5}M_\odot$ along with their corresponding assembly histories. The gray shaded areas represent the $68\%$ of the progenitor distribution, while the blue and red solid lines correspond to star-forming and quiescent galaxies of the same masses, with quiescent fractions taken from \citet{Moustakas+2013}. Figure \ref{fig:generalization_to_the_method} shows that star-forming galaxies have distinct median mass assembly histories compared to quiescent galaxies of the same final stellar mass $M_{\ast,0}$ \citep[see e.g.,][]{Clauwens+2016,Wang+2023}. Additionally, it shows that the median closely follows the progenitor that dominates the population. In general, this demonstrates how to sample a wide range of galaxy assembly histories using the approach described here, across a continuous range of galaxy stellar masses.

\begin{figure}
		\includegraphics[width=\columnwidth]{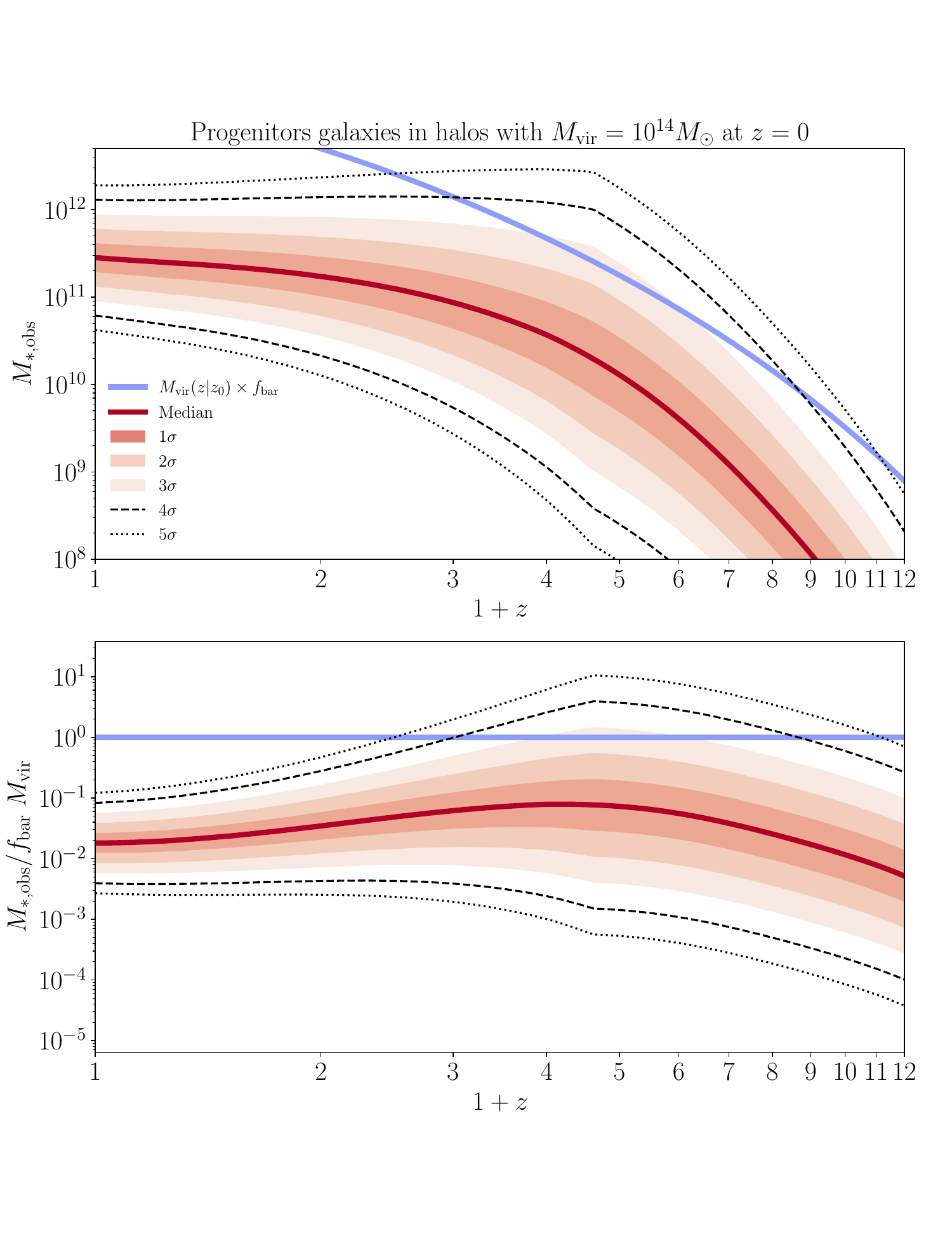}
			\caption{Median evolutionary tracks of galaxies within haloes with final masses of $M_{\rm vir}= 10^{14} M_\odot$ at $z=0$. The upper panel corresponds to the stellar mass assembly while the bottom panel is for the $M_\ast/f_{\rm bar}M_\text{vir}$ ratio. The various shaded areas indicate the 1 to $5\sigma_\text{tot}$ range of the distribution of galaxies with observed stellar masses, that is when combining in quadrature intrinsic scatter of the stellar-to-halo mass relation and random errors from stellar masses. The blue line indicates the universal baryon fraction. Observe that some galaxies are predicted to have mass ratios or masses bigger than the ones the universal baryon fraction allows. However, this occurs at the $\sim3-4\sigma_\text{tot}$ level.
 		}
	\label{fig:imposible_galaxies_2}
\end{figure}

\begin{figure}
		\includegraphics[width=\columnwidth]{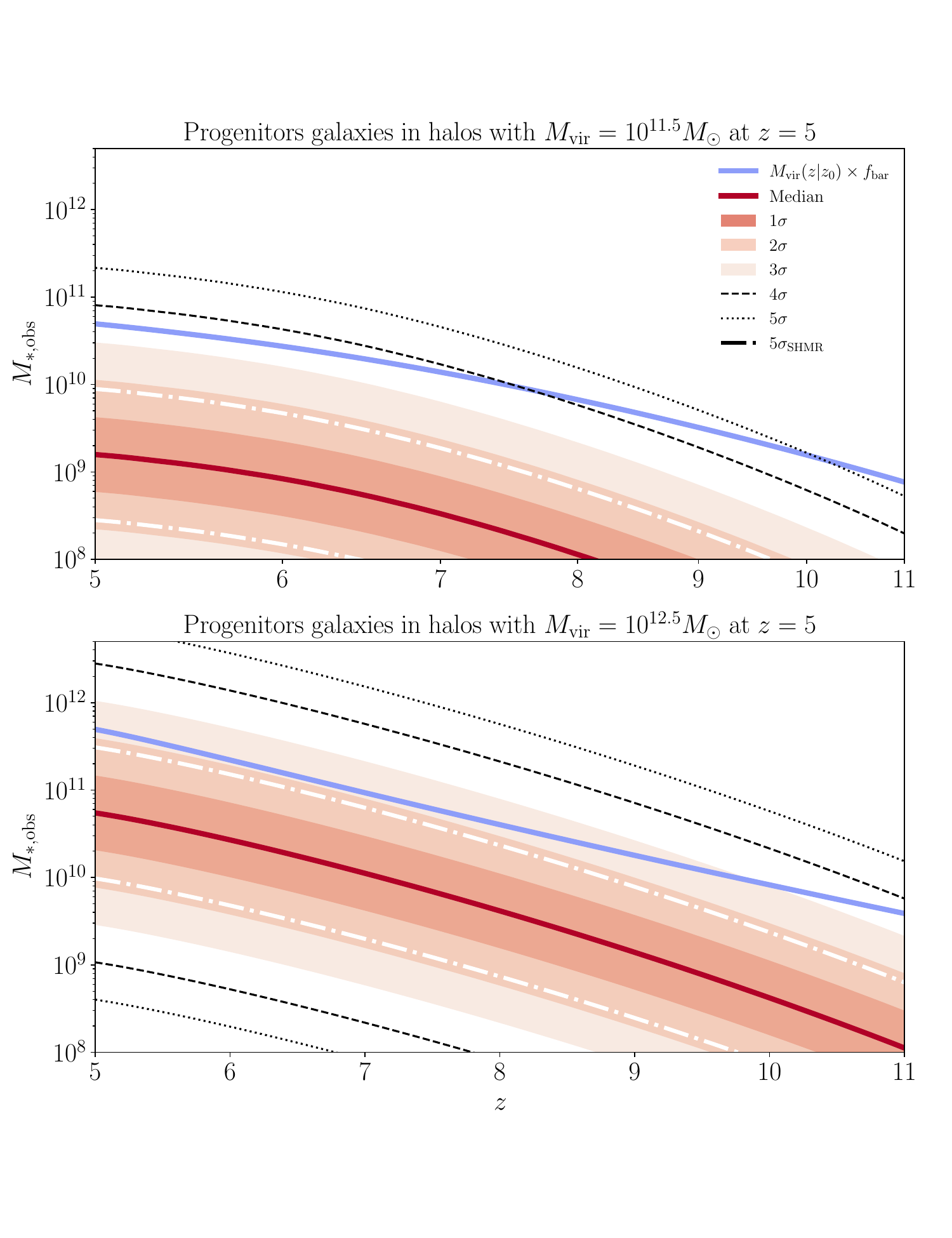}
			\caption{Similar to the upper panel of Figure \ref{fig:imposible_galaxies_2}, but for galaxies in haloes with $M_{\rm vir} = 10^{11.5} M_\odot$ and $M_{\rm vir} = 10^{12.5} M_\odot$ at $z=5$. For the low mass halo, galaxies at the $\sim3-4\sigma_\text{tot}$ level exceed the universal baryon fraction. For the high mass halo, galaxies at the $\sim2\sigma_\text{tot}$ level are already exceeding the universal baryon fraction. In the case of perfect mass measurements, in which case only the scatter around the stellar-to-halo mass relation accounts, we found that at the $5\sigma_\text{SHMR}$ level the possibility of impossible early galaxies is excluded as indicated by the dot-dashed lines. 
 		}
	\label{fig:imposible_galaxies_3}
\end{figure}

\subsection{Is there such a thing as an impossible galaxy?}
\label{sec:jwst_massive_gals}

As an example of how our approach can be applied to infer galaxy mass evolution from GSMFs reported at different redshifts, we revisit the so-called `impossible galaxies' problem. The James Webb Space Telescope (JWST) has pushed the boundaries of our understanding of the early universe. Initial JWST results suggested a potential overabundance of massive galaxies at very high redshifts, which appeared to be in tension with the standard $\Lambda$CDM cosmology \citep{Boylan-Kolchin_2023,Labbe+2023} or requiring extraordinary conversion of baryons into stars \citep{Dekel+2023,Xiao+2023,Zhaozhou+2024} leading to the impossible galaxies problem. Based on previous observations by the Hubble Space Telescope, some authors also pointed out this potential tension \citep[see e.g.,][]{Steinhardt+2016,Mason+2023}. However, this tension has diminished with the acquisition of spectroscopic data, which has improved redshift and mass estimates, and in some cases, indicated the presence of AGNs \citep[see e.g.,][]{Kocevski+2023,Napolitano+2024}. In this section, we do not aim to provide another solution to the impossible galaxies problem, but rather to explore the implications of our semi-empirical approach at high redshifts, and particularly understanding the impact of random errors at such early epochs \citep[see also ][]{Behroozi-Silk2018,Chen+2023}.

Figure \ref{fig:imposible_galaxies_2} illustrates the assembly history of galaxies within dark matter haloes with a final mass of $M_\text{vir} = 10^{14} M_\odot$ at $z=0$ (upper panel) and its corresponding stellar-to-baryonic halo mass ratio (lower panel). The solid line represents the median evolutionary track, while the various shaded areas indicate the 1 to $3\sigma$ range and the dashed and dotted black curves the 4 and $5\sigma$, respectively, of the distribution based on observed random errors, see Figure \ref{fig:ran_ms}, combined in quadrature with the intrinsic scatter of $\sigma_\text{SHMR} = 0.15$ dex around the stellar-to-halo mass relation. In other words, $\sigma_\text{tot}^2 = \sigma_{\ast,\text{ran}}^2(z) + \sigma_\text{SHMR}^2$. The blue line in the upper panel represents the universal baryon fraction times the virial mass at some redshift, while in the lower panel, it refers to the universal baryon fraction alone. This figure clearly shows that some observed galaxies will have stellar masses exceeding the universal baryon fraction, particularly those with observed stellar masses that are $\sim3-4\sigma_\text{tot}$ above the median at $z\sim2-8$. That is, less than $\sim0.2\%$ of the highest massive galaxies at these redshifts are expected to have observed stellar mass fractions exceeding the universal baryon fraction. Nonetheless, we emphasize that the distribution of observed stellar masses, $M_{\ast,\text{obs}}$, at those redshifts is composed of the intrinsic scatter around the stellar-to-halo mass relation, $\sigma_\text{SHMR} = 0.15$ dex, and the errors at high redshift $\sigma_{\ast,\text{ran}}= 0.4$. It follows then that the $1\sigma_\text{tot}$ contour at higher redshifts, corresponding to a value of $\sigma_\text{tot} \sim 0.43$ dex, is equivalent to $\sim3\sigma_\text{SHMR}$, which excludes galaxies with intrinsic stellar mass fractions (i.e., with perfect mass measurements) that are higher than the baryon fraction at all $z$. In other words, intrinsic stellar masses do not exceed maximal masses given by the universal baryon fraction. In our context, the observation of stellar masses higher than these extreme masses at higher redshifts is thus primarily due to random variations in stellar mass distribution. We note that \citet{Behroozi-Silk2018} and more recently \citet{Chen+2023} reached a similar conclusion.

We provide two more examples. Figure \ref{fig:imposible_galaxies_3} is the same as the top panel of Figure \ref{fig:imposible_galaxies_2}, but this time it illustrates the assembly histories in two different host haloes: one with a final mass of $M_\text{vir} = 10^{11.5} M_\odot$ and the other with a final mass of $M_\text{vir} = 10^{12.5} M_\odot$, both at $z=5$. For the smaller halo, from $z=5$ to $z\sim8$, we observe that galaxies with stellar masses around $\sim3-4\sigma_\text{tot}$ exceed the baryon fraction. At redshifts greater than $z\sim9$, only galaxies around $5\sigma_\text{tot}$ exceed the baryon fraction. For the halo with a final mass of $M_\text{vir} = 10^{12.5} M_\odot$, galaxies around $2\sigma_\text{tot}$ exceed the universal baryon fraction within the redshift range of $z\sim5-8$, that is $\sim2\%$ of this population at any redshift. At higher redshifts, this occurs between $3-4\sigma_\text{tot}$. Once again, galaxies that exceed the baryon fraction are likely due to random errors affecting their observed values. To aid interpretation, it is important to note that if we were to trace the intrinsic distribution around the stellar-to-halo mass relation, our shaded areas of $2\sigma_\text{tot}$ would correspond to approximately $6\sigma_\text{SHMR}$, which \emph{strongly excludes} the possibility of ``impossible early galaxies'' in the case of perfect mass measurements. However, this does allow for the possibility of some galaxies having stellar masses close to the maximal masses given by the universal baryon fraction, as shown by the $5\sigma_\mathrm{SHMR}$ white dot-dashed line.

The empirical evidence in this section suggests that some galaxies are expected to have stellar masses exceeding those predicted by the universal baryon fraction. These are extreme cases, where galaxies with stellar masses $3-5\sigma_\text{tot}$ from the median are too massive to be easily explained by the $\Lambda$CDM model. However, it is not a coincidence that this tension appears to be more pronounced at higher redshifts, where random errors increase significantly, reaching $\sigma_{\ast,\text{ran}} = 0.4$ dex at redshifts $z > 4$. In contrast, the intrinsic scatter around the stellar-to-halo mass relation, 0.15 dex, would predict galaxies closer to, but not exceeding, the universal baryon fraction under ideal measurement conditions. This highlights the importance of understanding random errors in stellar mass measurements to assess whether there is a genuine tension between observed massive galaxies and our standard model of galaxy formation and evolution.

\section{Conclusions}
\label{sec:conclusions}

In this paper, we revisited the cumulative number density matching approach \citep{vanDokkum+2010} used to infer the mass growth of galaxies from the GSMF at different redshifts, focusing on the perspective of evolving halo cumulative number density as implied by the median growth of dark matter haloes and using a semi-empirical galaxy-halo connection approach. We also discussed how this approach can be easily extended to account for different galaxy types. As an example of the application of our approach, we have derived the evolutionary tracks of galaxies to investigate whether the concept of ``impossible galaxies'' is plausible in light of the recent claims of massive galaxies at high redshifts \citep[$z\sim6-10$, see e.g.,][]{Labbe+2023}, which suggest star formation efficiencies exceeding those predicted by the universal baryon fraction \citep{Boylan-Kolchin_2023}. Our results and conclusions are as follows:

\begin{itemize}

    \item Random errors affecting the measurement of stellar masses increase from approximately $0.1$ dex at low redshifts ($z\lesssim0.4$) to around $0.4$ dex at higher redshifts ($z\gtrsim 4$), as shown in Figure \ref{fig:ran_ms}. 

    \item For the empirical cumulative number density matching approach, accounting for random errors in stellar mass leads to smaller inferred stellar masses at a fixed redshift for a given progenitor. It also results in increased observed cumulative number densities (i.e., convolved with errors) at higher redshifts, Figure \ref{fig:numb_density_obs_vs_deconv}.

    \item Using $N$-body cosmological simulations, we compute the median assembly history of dark matter haloes and track their cumulative number density across redshifts. We found that halo number densities are far from constant, and depend on mass and redshift. On average, they exhibit an slope of approximately $0.2$ dex per $\Delta z$ , consistent with the increase of 0.16 dex per $\Delta z$ suggested by \citet{Behroozi+2013f}, Figure  \ref{fig:halo_numb_density}. At higher redshifts, our results suggest that the constant cumulative number density assumption becomes more accurate, Figure \ref{fig:halo_numb_density_at_z2&z4} , which is in tension with the results from \citet{Behroozi+2013f}.

    \item Assuming that galaxies form and evolve within dark matter haloes, we use evolving halo cumulative number densities and the subhalo abundance matching technique to infer realistic median evolutionary tracks of galaxies. Compared to tracks based on the empirical cumulative number density matching approach, which accounts for random errors in stellar masses, the evolving halo cumulative number density approach leads to galaxies that are a factor of $\sim2-3$ times smaller in mass at $z\sim2-4$. At higher redshifts, we found that the differences are more pronounced, with masses inferred by evolving halo number densities being smaller by about an order of magnitude at $z\sim7$, Figure \ref{fig:galaxy_tracks_from_halo_num_tracks}.

    \item We discuss how our approach can be extended to incorporate a segregation of the halo assembly histories by a second property using a simple analytical model. From this exercise, connecting halo concentration with the star formation activity of galaxies, we show that star-forming and quiescent galaxies with the same final stellar mass have distinct median progenitors. Quiescent galaxies assemble earlier than star-forming ones, with the median model being closer to the dominant population (see Figure \ref{fig:generalization_to_the_method}).

    \item We use the evolving halo cumulative number density approach to investigate the evolutionary tracks of galaxies at different redshifts, focusing on the early galaxy problem. When accounting both for random errors in the stellar mass determination and the scatter around the stellar-to-halo mass relation, $\sigma_\text{tot}^2 = \sigma_{\ast,\text{ran}}^2(z) + \sigma_\text{SHMR}^2$, 
    we find that galaxy progenitors within a halo with final mass of $M_\text{vir} \sim 10^{14} M_\odot$ at $z=0$, which are $\sim3-4\sigma_\text{tot}$, at $z\sim2-8$ have stellar masses exceeding the limit imposed by the universal baryon fraction, Figure \ref{fig:imposible_galaxies_2}. 
    The issue becomes more pronounced when studying galaxy progenitors at $z=5$. For $M_\text{vir} = 10^{11.5} M_\odot$ at $z=5$, galaxy progenitors that are $\sim3-4\sigma_\text{tot}$ exceed the baryon fraction at $z\sim5-8$, while for $M_\text{vir} = 10^{12.5} M_\odot$, this happens for galaxies that are $\sim2\sigma_\text{tot}$, Figure \ref{fig:imposible_galaxies_3}.

    \item We note, however, that if the scatter around the stellar-to-halo mass relation remains constant with redshift and is of the order of $\sigma_\text{SHMR}\sim0.15$ dex, then in the ideal scenario of perfect mass measurements, the early galaxy formation problem is ruled out at a significance level of $\sim5-6\sigma$, Figure \ref{fig:imposible_galaxies_3}.
    
\end{itemize}

We have shown the inaccuracies of assuming a constant cumulative galaxy number density for the progenitors of a given galaxy and how this leads to an overestimation of stellar masses at high redshifts using the empirical number density matching technique. Our results, which utilize the galaxy-halo connection and account for random errors in stellar mass determination, can be readily applied to correct and improve the estimation of the average stellar mass growth of galaxies using this technique.

Finally, stellar mass measurements are now routinely performed when the SED of a galaxy is analyzed. However, errors in these mass estimates have been explored very little in the past, see Figure \ref{fig:ran_ms}. Our results clearly show how random errors impact inferences on galaxy formation and evolution, whether using purely empirical approaches like the cumulative number density matching approach, semi-empirical methods such as the one developed here, or more complex models that utilize the full merger trees of dark matter haloes \citep[e.g.,][]{Behroozi+2019}. We propose that understanding random errors in stellar masses will be key to alleviating the tension in the early galaxy formation problem and understanding the origin of the scatter around the stellar-to-halo mass relation. Finally, we also note that the Gaussian errors assumed here may represent an oversimplification. 

\section*{Acknowledgements}

We thank the referee, Kai Wang, for the helpful comments that have helped to improve the presentation of this paper. We are grateful for helpful conversations with Brant Robertson and Peter Behroozi. ARP and VAR acknowledge financial support from DGAPA-PAPIIT grants IN106823, IN106124 and IN106924 and from CONAHCyT grant CF G-543. We thank contributors to the \textsc{Python} programming language\footnote{\url{https://www.python.org/}}, \textsc{SciPy}\footnote{\url{https://www.scipy.org/} \citep{2020SciPy-NMeth}}, \textsc{NumPy}\footnote{\url{https://numpy.org/} \citep{2020NumPy-Array}}, \textsc{Matplotlib}\footnote{\url{https://matplotlib.org/} \citep{Hunter:2007}}, and the free and open-source community.

\section*{Data Availability}

A publicly available code for evaluating semi-empirical galaxy evolutionary tracks, as described in Sections \ref{sec:gal_evol_tracks} and \ref{sec:generaliztion_of_the_approach}, can be found on \href{https://github.com/TheConCHaProject/ConCHa}{GitHub link \faGithub}, or via email at \href{mailto:galhalo\_conn@astro.unam.mx}{galhalo\_conn@astro.unam.mx}.



\bibliographystyle{mnras}




\bibliography{Bibliography} 

\begin{thebibliography}{}
\makeatletter
\relax
\def\mn@urlcharsother{\let\do\@makeother \do\$\do\&\do\#\do\^\do\_\do\%\do\~}
\def\mn@doi{\begingroup\mn@urlcharsother \@ifnextchar [ {\mn@doi@}
  {\mn@doi@[]}}
\def\mn@doi@[#1]#2{\def\@tempa{#1}\ifx\@tempa\@empty \href
  {http://dx.doi.org/#2} {doi:#2}\else \href {http://dx.doi.org/#2} {#1}\fi
  \endgroup}
\def\mn@eprint#1#2{\mn@eprint@#1:#2::\@nil}
\def\mn@eprint@arXiv#1{\href {http://arxiv.org/abs/#1} {{\tt arXiv:#1}}}
\def\mn@eprint@dblp#1{\href {http://dblp.uni-trier.de/rec/bibtex/#1.xml}
  {dblp:#1}}
\def\mn@eprint@#1:#2:#3:#4\@nil{\def\@tempa {#1}\def\@tempb {#2}\def\@tempc
  {#3}\ifx \@tempc \@empty \let \@tempc \@tempb \let \@tempb \@tempa \fi \ifx
  \@tempb \@empty \def\@tempb {arXiv}\fi \@ifundefined
  {mn@eprint@\@tempb}{\@tempb:\@tempc}{\expandafter \expandafter \csname
  mn@eprint@\@tempb\endcsname \expandafter{\@tempc}}}

\bibitem[\protect\citeauthoryear{{Behroozi} \& {Silk}}{{Behroozi} \&
  {Silk}}{2018}]{Behroozi-Silk2018}
{Behroozi} P.,  {Silk} J.,  2018, \mn@doi [\mnras] {10.1093/mnras/sty945},
  \href {https://ui.adsabs.harvard.edu/abs/2018MNRAS.477.5382B} {477, 5382}

\bibitem[\protect\citeauthoryear{{Behroozi}, {Conroy}  \&
  {Wechsler}}{{Behroozi} et~al.}{2010}]{Behroozi+2010}
{Behroozi} P.~S.,  {Conroy} C.,   {Wechsler} R.~H.,  2010, \mn@doi [\apj]
  {10.1088/0004-637X/717/1/379}, \href
  {http://adsabs.harvard.edu/abs/2010ApJ...717..379B} {717, 379}

\bibitem[\protect\citeauthoryear{{Behroozi}, {Wechsler}  \& {Wu}}{{Behroozi}
  et~al.}{2013a}]{Behroozi+2013d}
{Behroozi} P.~S.,  {Wechsler} R.~H.,   {Wu} H.-Y.,  2013a, \mn@doi [\apj]
  {10.1088/0004-637X/762/2/109}, \href
  {http://adsabs.harvard.edu/abs/2013ApJ...762..109B} {762, 109}

\bibitem[\protect\citeauthoryear{{Behroozi}, {Wechsler}, {Wu}, {Busha},
  {Klypin}  \& {Primack}}{{Behroozi} et~al.}{2013b}]{Behroozi+2013b}
{Behroozi} P.~S.,  {Wechsler} R.~H.,  {Wu} H.-Y.,  {Busha} M.~T.,  {Klypin}
  A.~A.,   {Primack} J.~R.,  2013b, \mn@doi [\apj]
  {10.1088/0004-637X/763/1/18}, \href
  {http://adsabs.harvard.edu/abs/2013ApJ...763...18B} {763, 18}

\bibitem[\protect\citeauthoryear{{Behroozi}, {Wechsler}  \&
  {Conroy}}{{Behroozi} et~al.}{2013c}]{Behroozi+2013}
{Behroozi} P.~S.,  {Wechsler} R.~H.,   {Conroy} C.,  2013c, \mn@doi [\apj]
  {10.1088/0004-637X/770/1/57}, \href
  {http://adsabs.harvard.edu/abs/2013ApJ...770...57B} {770, 57}

\bibitem[\protect\citeauthoryear{{Behroozi}, {Marchesini}, {Wechsler},
  {Muzzin}, {Papovich}  \& {Stefanon}}{{Behroozi}
  et~al.}{2013d}]{Behroozi+2013f}
{Behroozi} P.~S.,  {Marchesini} D.,  {Wechsler} R.~H.,  {Muzzin} A.,
  {Papovich} C.,   {Stefanon} M.,  2013d, \mn@doi [\apjl]
  {10.1088/2041-8205/777/1/L10}, \href
  {https://ui.adsabs.harvard.edu/abs/2013ApJ...777L..10B} {777, L10}

\bibitem[\protect\citeauthoryear{{Behroozi}, {Wechsler}, {Hearin}  \&
  {Conroy}}{{Behroozi} et~al.}{2019}]{Behroozi+2019}
{Behroozi} P.,  {Wechsler} R.~H.,  {Hearin} A.~P.,   {Conroy} C.,  2019,
  \mn@doi [\mnras] {10.1093/mnras/stz1182}, \href
  {https://ui.adsabs.harvard.edu/abs/2019MNRAS.488.3143B} {488, 3143}

\bibitem[\protect\citeauthoryear{{Boylan-Kolchin}}{{Boylan-Kolchin}}{2023}]{Boylan-Kolchin_2023}
{Boylan-Kolchin} M.,  2023, \mn@doi [Nature Astronomy]
  {10.1038/s41550-023-01937-7}, \href
  {https://ui.adsabs.harvard.edu/abs/2023NatAs...7..731B} {7, 731}

\bibitem[\protect\citeauthoryear{{Calette}, {Rodr{\'\i}guez-Puebla},
  {Avila-Reese}  \& {Lagos}}{{Calette} et~al.}{2021}]{Calette+2021b}
{Calette} A.~R.,  {Rodr{\'\i}guez-Puebla} A.,  {Avila-Reese} V.,   {Lagos} C.
  d.~P.,  2021, \mn@doi [\mnras] {10.1093/mnras/stab1788}, \href
  {https://ui.adsabs.harvard.edu/abs/2021MNRAS.506.1507C} {506, 1507}

\bibitem[\protect\citeauthoryear{{Cattaneo}, {Dekel}, {Faber}  \&
  {Guiderdoni}}{{Cattaneo} et~al.}{2008}]{Cattaneo+2008}
{Cattaneo} A.,  {Dekel} A.,  {Faber} S.~M.,   {Guiderdoni} B.,  2008, \mn@doi
  [\mnras] {10.1111/j.1365-2966.2008.13562.x}, \href
  {http://adsabs.harvard.edu/abs/2008MNRAS.389..567C} {389, 567}

\bibitem[\protect\citeauthoryear{{Chabrier}}{{Chabrier}}{2003}]{Chabrier2003}
{Chabrier} G.,  2003, \mn@doi [\pasp] {10.1086/376392}, \href
  {http://adsabs.harvard.edu/abs/2003PASP..115..763C} {115, 763}

\bibitem[\protect\citeauthoryear{{Chen}, {Mo}  \& {Wang}}{{Chen}
  et~al.}{2023}]{Chen+2023}
{Chen} Y.,  {Mo} H.~J.,   {Wang} K.,  2023, \mn@doi [\mnras]
  {10.1093/mnras/stad2866}, \href
  {https://ui.adsabs.harvard.edu/abs/2023MNRAS.526.2542C} {526, 2542}

\bibitem[\protect\citeauthoryear{{Clauwens}, {Franx}  \& {Schaye}}{{Clauwens}
  et~al.}{2016}]{Clauwens+2016}
{Clauwens} B.,  {Franx} M.,   {Schaye} J.,  2016, \mn@doi [\mnras]
  {10.1093/mnrasl/slw137}, \href
  {https://ui.adsabs.harvard.edu/abs/2016MNRAS.463L...1C} {463, L1}

\bibitem[\protect\citeauthoryear{{Conroy}}{{Conroy}}{2013}]{Conroy2013}
{Conroy} C.,  2013, \mn@doi [\araa] {10.1146/annurev-astro-082812-141017},
  \href {http://adsabs.harvard.edu/abs/2013ARA%26A..51..393C} {51, 393}

\bibitem[\protect\citeauthoryear{{Conroy} \& {Wechsler}}{{Conroy} \&
  {Wechsler}}{2009}]{Conroy+2009}
{Conroy} C.,  {Wechsler} R.~H.,  2009, \mn@doi [\apj]
  {10.1088/0004-637X/696/1/620}, \href
  {http://adsabs.harvard.edu/abs/2009ApJ...696..620C} {696, 620}

\bibitem[\protect\citeauthoryear{{Conroy}, {Wechsler}  \& {Kravtsov}}{{Conroy}
  et~al.}{2006}]{Conroy+2006}
{Conroy} C.,  {Wechsler} R.~H.,   {Kravtsov} A.~V.,  2006, \mn@doi [\apj]
  {10.1086/503602}, \href {http://adsabs.harvard.edu/abs/2006ApJ...647..201C}
  {647, 201}

\bibitem[\protect\citeauthoryear{{Conroy}, {Gunn}  \& {White}}{{Conroy}
  et~al.}{2009}]{Conroy+2009a}
{Conroy} C.,  {Gunn} J.~E.,   {White} M.,  2009, \mn@doi [\apj]
  {10.1088/0004-637X/699/1/486}, \href
  {https://ui.adsabs.harvard.edu/abs/2009ApJ...699..486C} {699, 486}

\bibitem[\protect\citeauthoryear{{Conselice}, {Wilkinson}, {Duncan}  \&
  {Mortlock}}{{Conselice} et~al.}{2016}]{Conselice+2016}
{Conselice} C.~J.,  {Wilkinson} A.,  {Duncan} K.,   {Mortlock} A.,  2016,
  \mn@doi [\apj] {10.3847/0004-637X/830/2/83}, \href
  {http://adsabs.harvard.edu/abs/2016ApJ...830...83C} {830, 83}

\bibitem[\protect\citeauthoryear{{Correa} \& {Schaye}}{{Correa} \&
  {Schaye}}{2020}]{Correa_Schaye2020}
{Correa} C.~A.,  {Schaye} J.,  2020, \mn@doi [\mnras] {10.1093/mnras/staa3053},
  \href {https://ui.adsabs.harvard.edu/abs/2020MNRAS.499.3578C} {499, 3578}

\bibitem[\protect\citeauthoryear{{Crain} et~al.,}{{Crain}
  et~al.}{2015}]{Crain+2015}
{Crain} R.~A.,  et~al., 2015, \mn@doi [\mnras] {10.1093/mnras/stv725}, \href
  {https://ui.adsabs.harvard.edu/abs/2015MNRAS.450.1937C} {450, 1937}

\bibitem[\protect\citeauthoryear{{Danieli}, {Greene}, {Carlsten}, {Jiang},
  {Beaton}  \& {Goulding}}{{Danieli} et~al.}{2023}]{Danieli+2023}
{Danieli} S.,  {Greene} J.~E.,  {Carlsten} S.,  {Jiang} F.,  {Beaton} R.,
  {Goulding} A.~D.,  2023, \mn@doi [\apj] {10.3847/1538-4357/acefbd}, \href
  {https://ui.adsabs.harvard.edu/abs/2023ApJ...956....6D} {956, 6}

\bibitem[\protect\citeauthoryear{{Davidzon} et~al.,}{{Davidzon}
  et~al.}{2017}]{Davidzon+2017}
{Davidzon} I.,  et~al., 2017, \mn@doi [\aap] {10.1051/0004-6361/201730419},
  \href {https://ui.adsabs.harvard.edu/abs/2017A&A...605A..70D} {605, A70}

\bibitem[\protect\citeauthoryear{{Dekel} \& {Mandelker}}{{Dekel} \&
  {Mandelker}}{2014}]{Dekel+2014}
{Dekel} A.,  {Mandelker} N.,  2014, \mn@doi [\mnras] {10.1093/mnras/stu1427},
  \href {http://adsabs.harvard.edu/abs/2014MNRAS.444.2071D} {444, 2071}

\bibitem[\protect\citeauthoryear{{Dekel}, {Sarkar}, {Birnboim}, {Mandelker}  \&
  {Li}}{{Dekel} et~al.}{2023}]{Dekel+2023}
{Dekel} A.,  {Sarkar} K.~C.,  {Birnboim} Y.,  {Mandelker} N.,   {Li} Z.,  2023,
  \mn@doi [\mnras] {10.1093/mnras/stad1557}, \href
  {https://ui.adsabs.harvard.edu/abs/2023MNRAS.523.3201D} {523, 3201}

\bibitem[\protect\citeauthoryear{{Diemer} \& {Kravtsov}}{{Diemer} \&
  {Kravtsov}}{2015}]{Diemer_Kravtsov2015}
{Diemer} B.,  {Kravtsov} A.~V.,  2015, \mn@doi [\apj]
  {10.1088/0004-637X/799/1/108}, \href
  {https://ui.adsabs.harvard.edu/abs/2015ApJ...799..108D} {799, 108}

\bibitem[\protect\citeauthoryear{{Dragomir}, {Rodr{\'{\i}}guez-Puebla},
  {Primack}  \& {Lee}}{{Dragomir} et~al.}{2018}]{Dragomir+2018}
{Dragomir} R.,  {Rodr{\'{\i}}guez-Puebla} A.,  {Primack} J.~R.,   {Lee} C.~T.,
  2018, \mn@doi [\mnras] {10.1093/mnras/sty283}, \href
  {http://adsabs.harvard.edu/abs/2018MNRAS.476..741D} {476, 741}

\bibitem[\protect\citeauthoryear{{Drory} \& {Alvarez}}{{Drory} \&
  {Alvarez}}{2008}]{DroryAlvarez2008}
{Drory} N.,  {Alvarez} M.,  2008, \mn@doi [\apj] {10.1086/588006}, \href
  {http://adsabs.harvard.edu/abs/2008ApJ...680...41D} {680, 41}

\bibitem[\protect\citeauthoryear{{Dutton} \& {Macci{\`o}}}{{Dutton} \&
  {Macci{\`o}}}{2014}]{Dutton+2014}
{Dutton} A.~A.,  {Macci{\`o}} A.~V.,  2014, \mn@doi [\mnras]
  {10.1093/mnras/stu742}, \href
  {https://ui.adsabs.harvard.edu/abs/2014MNRAS.441.3359D} {441, 3359}

\bibitem[\protect\citeauthoryear{{Eddington}}{{Eddington}}{1913}]{Eddington1913}
{Eddington} A.~S.,  1913, \mn@doi [\mnras] {10.1093/mnras/73.5.359}, \href
  {https://ui.adsabs.harvard.edu/abs/1913MNRAS..73..359E} {73, 359}

\bibitem[\protect\citeauthoryear{{Eddington}}{{Eddington}}{1940}]{Eddington1940}
{Eddington} Sir A.~S.,  1940, \mn@doi [\mnras] {10.1093/mnras/100.5.354}, \href
  {http://adsabs.harvard.edu/abs/1940MNRAS.100..354E} {100, 354}

\bibitem[\protect\citeauthoryear{{Firmani} \& {Avila-Reese}}{{Firmani} \&
  {Avila-Reese}}{2010}]{Firmani+2010a}
{Firmani} C.,  {Avila-Reese} V.,  2010, \mn@doi [\apj]
  {10.1088/0004-637X/723/1/755}, \href
  {http://adsabs.harvard.edu/abs/2010ApJ...723..755F} {723, 755}

\bibitem[\protect\citeauthoryear{{Frenk}, {White}, {Davis}  \&
  {Efstathiou}}{{Frenk} et~al.}{1988}]{Frenk+1988}
{Frenk} C.~S.,  {White} S. D.~M.,  {Davis} M.,   {Efstathiou} G.,  1988,
  \mn@doi [\apj] {10.1086/166213}, \href
  {https://ui.adsabs.harvard.edu/abs/1988ApJ...327..507F} {327, 507}

\bibitem[\protect\citeauthoryear{{Fu} et~al.,}{{Fu} et~al.}{2024}]{Fu+2024}
{Fu} H.,  et~al., 2024, \mn@doi [\mnras] {10.1093/mnras/stae1492}, \href
  {https://ui.adsabs.harvard.edu/abs/2024MNRAS.532..177F} {532, 177}

\bibitem[\protect\citeauthoryear{{Grazian} et~al.,}{{Grazian}
  et~al.}{2015}]{Grazian+2015}
{Grazian} A.,  et~al., 2015, \mn@doi [\aap] {10.1051/0004-6361/201424750},
  \href {http://adsabs.harvard.edu/abs/2015A%26A...575A..96G} {575, A96}

\bibitem[\protect\citeauthoryear{{Grylls}, {Shankar}, {Leja}, {Menci},
  {Moster}, {Behroozi}  \& {Zanisi}}{{Grylls} et~al.}{2020}]{Grylls+2020}
{Grylls} P.~J.,  {Shankar} F.,  {Leja} J.,  {Menci} N.,  {Moster} B.,
  {Behroozi} P.,   {Zanisi} L.,  2020, \mn@doi [\mnras]
  {10.1093/mnras/stz2956}, \href
  {https://ui.adsabs.harvard.edu/abs/2020MNRAS.491..634G} {491, 634}

\bibitem[\protect\citeauthoryear{Harris et~al.,}{Harris
  et~al.}{2020}]{2020NumPy-Array}
Harris C.~R.,  et~al., 2020, \mn@doi [Nature] {10.1038/s41586-020-2649-2}, 585,
  357–362

\bibitem[\protect\citeauthoryear{{Harvey} et~al.,}{{Harvey}
  et~al.}{2024}]{Harvey+2024}
{Harvey} T.,  et~al., 2024, \mn@doi [arXiv e-prints]
  {10.48550/arXiv.2403.03908}, \href
  {https://ui.adsabs.harvard.edu/abs/2024arXiv240303908H} {p. arXiv:2403.03908}

\bibitem[\protect\citeauthoryear{{Hearin} \& {Watson}}{{Hearin} \&
  {Watson}}{2013a}]{HearinWatson2013}
{Hearin} A.~P.,  {Watson} D.~F.,  2013a, \mn@doi [\mnras]
  {10.1093/mnras/stt1374}, \href
  {http://adsabs.harvard.edu/abs/2013MNRAS.435.1313H} {435, 1313}

\bibitem[\protect\citeauthoryear{{Hearin} \& {Watson}}{{Hearin} \&
  {Watson}}{2013b}]{Hearin_Watson2013}
{Hearin} A.~P.,  {Watson} D.~F.,  2013b, \mn@doi [\mnras]
  {10.1093/mnras/stt1374}, \href
  {https://ui.adsabs.harvard.edu/abs/2013MNRAS.435.1313H} {435, 1313}

\bibitem[\protect\citeauthoryear{Hunter}{Hunter}{2007}]{Hunter:2007}
Hunter J.~D.,  2007, \mn@doi [Computing in Science \& Engineering]
  {10.1109/MCSE.2007.55}, 9, 90

\bibitem[\protect\citeauthoryear{{Ilbert} et~al.,}{{Ilbert}
  et~al.}{2013}]{Ilbert+2013}
{Ilbert} O.,  et~al., 2013, \mn@doi [\aap] {10.1051/0004-6361/201321100}, \href
  {http://adsabs.harvard.edu/abs/2013A%26A...556A..55I} {556, A55}

\bibitem[\protect\citeauthoryear{{Kajisawa} et~al.,}{{Kajisawa}
  et~al.}{2009}]{Kajisawa+2009}
{Kajisawa} M.,  et~al., 2009, \mn@doi [\apj] {10.1088/0004-637X/702/2/1393},
  \href {http://adsabs.harvard.edu/abs/2009ApJ...702.1393K} {702, 1393}

\bibitem[\protect\citeauthoryear{{Kakos}, {Rodriguez-Puebla}, {Primack},
  {Faber}, {Koo}, {Behroozi}  \& {Avila-Reese}}{{Kakos}
  et~al.}{2024}]{Kakos+2024}
{Kakos} J.,  {Rodriguez-Puebla} A.,  {Primack} J.~R.,  {Faber} S.~M.,  {Koo}
  D.~C.,  {Behroozi} P.,   {Avila-Reese} V.,  2024, arXiv e-prints, \href
  {https://ui.adsabs.harvard.edu/abs/2024arXiv240301393K} {p. arXiv:2403.01393}

\bibitem[\protect\citeauthoryear{{Kauffmann} et~al.,}{{Kauffmann}
  et~al.}{2003}]{Kauffmann+2003}
{Kauffmann} G.,  et~al., 2003, \mn@doi [\mnras]
  {10.1046/j.1365-8711.2003.06291.x}, \href
  {http://adsabs.harvard.edu/abs/2003MNRAS.341...33K} {341, 33}

\bibitem[\protect\citeauthoryear{{Kawinwanichakij} et~al.,}{{Kawinwanichakij}
  et~al.}{2021}]{Kawinwanichakij+2021}
{Kawinwanichakij} L.,  et~al., 2021, \mn@doi [\apj] {10.3847/1538-4357/ac1f21},
  \href {https://ui.adsabs.harvard.edu/abs/2021ApJ...921...38K} {921, 38}

\bibitem[\protect\citeauthoryear{{Klypin}, {Yepes}, {Gottl{\"o}ber}, {Prada}
  \& {He{\ss}}}{{Klypin} et~al.}{2016}]{Klypin+2016}
{Klypin} A.,  {Yepes} G.,  {Gottl{\"o}ber} S.,  {Prada} F.,   {He{\ss}} S.,
  2016, \mn@doi [\mnras] {10.1093/mnras/stw248}, \href
  {https://ui.adsabs.harvard.edu/abs/2016MNRAS.457.4340K} {457, 4340}

\bibitem[\protect\citeauthoryear{{Kocevski} et~al.,}{{Kocevski}
  et~al.}{2023}]{Kocevski+2023}
{Kocevski} D.~D.,  et~al., 2023, \mn@doi [\apjl] {10.3847/2041-8213/ace5a0},
  \href {https://ui.adsabs.harvard.edu/abs/2023ApJ...954L...4K} {954, L4}

\bibitem[\protect\citeauthoryear{{Labb{\'e}} et~al.,}{{Labb{\'e}}
  et~al.}{2023}]{Labbe+2023}
{Labb{\'e}} I.,  et~al., 2023, \mn@doi [\nat] {10.1038/s41586-023-05786-2},
  \href {https://ui.adsabs.harvard.edu/abs/2023Natur.616..266L} {616, 266}

\bibitem[\protect\citeauthoryear{{Leja}, {van Dokkum}  \& {Franx}}{{Leja}
  et~al.}{2013}]{Leja+2013}
{Leja} J.,  {van Dokkum} P.,   {Franx} M.,  2013, \mn@doi [\apj]
  {10.1088/0004-637X/766/1/33}, \href
  {https://ui.adsabs.harvard.edu/abs/2013ApJ...766...33L} {766, 33}

\bibitem[\protect\citeauthoryear{{Leja}, {Speagle}, {Johnson}, {Conroy}, {van
  Dokkum}  \& {Franx}}{{Leja} et~al.}{2020}]{Leja+2020}
{Leja} J.,  {Speagle} J.~S.,  {Johnson} B.~D.,  {Conroy} C.,  {van Dokkum} P.,
   {Franx} M.,  2020, \mn@doi [\apj] {10.3847/1538-4357/ab7e27}, \href
  {https://ui.adsabs.harvard.edu/abs/2020ApJ...893..111L} {893, 111}

\bibitem[\protect\citeauthoryear{{Li}, {Dekel}, {Sarkar}, {Aung}, {Giavalisco},
  {Mandelker}  \& {Tacchella}}{{Li} et~al.}{2024}]{Zhaozhou+2024}
{Li} Z.,  {Dekel} A.,  {Sarkar} K.~C.,  {Aung} H.,  {Giavalisco} M.,
  {Mandelker} N.,   {Tacchella} S.,  2024, \mn@doi [\aap]
  {10.1051/0004-6361/202348727}, \href
  {https://ui.adsabs.harvard.edu/abs/2024A&A...690A.108L} {690, A108}

\bibitem[\protect\citeauthoryear{{Loeb} \& {Peebles}}{{Loeb} \&
  {Peebles}}{2003}]{LoebPeebles2003}
{Loeb} A.,  {Peebles} P.~J.~E.,  2003, \mn@doi [\apj] {10.1086/374349}, \href
  {https://ui.adsabs.harvard.edu/abs/2003ApJ...589...29L} {589, 29}

\bibitem[\protect\citeauthoryear{{Mandelbaum}, {Seljak}, {Kauffmann}, {Hirata}
  \& {Brinkmann}}{{Mandelbaum} et~al.}{2006}]{Mandelbaum+2006}
{Mandelbaum} R.,  {Seljak} U.,  {Kauffmann} G.,  {Hirata} C.~M.,   {Brinkmann}
  J.,  2006, \mn@doi [\mnras] {10.1111/j.1365-2966.2006.10156.x}, \href
  {http://adsabs.harvard.edu/abs/2006MNRAS.368..715M} {368, 715}

\bibitem[\protect\citeauthoryear{{Mandelbaum}, {Wang}, {Zu}, {White},
  {Henriques}  \& {More}}{{Mandelbaum} et~al.}{2016}]{Mandelbaum+2016}
{Mandelbaum} R.,  {Wang} W.,  {Zu} Y.,  {White} S.,  {Henriques} B.,   {More}
  S.,  2016, \mn@doi [\mnras] {10.1093/mnras/stw188}, \href
  {https://ui.adsabs.harvard.edu/abs/2016MNRAS.457.3200M} {457, 3200}

\bibitem[\protect\citeauthoryear{{Masaki}, {Lin}  \& {Yoshida}}{{Masaki}
  et~al.}{2013}]{Masaki+2013}
{Masaki} S.,  {Lin} Y.-T.,   {Yoshida} N.,  2013, \mn@doi [\mnras]
  {10.1093/mnras/stt1729}, \href
  {http://adsabs.harvard.edu/abs/2013MNRAS.436.2286M} {436, 2286}

\bibitem[\protect\citeauthoryear{{Mason}, {Trenti}  \& {Treu}}{{Mason}
  et~al.}{2023}]{Mason+2023}
{Mason} C.~A.,  {Trenti} M.,   {Treu} T.,  2023, \mn@doi [\mnras]
  {10.1093/mnras/stad035}, \href
  {https://ui.adsabs.harvard.edu/abs/2023MNRAS.521..497M} {521, 497}

\bibitem[\protect\citeauthoryear{{Mitra}, {Dav{\'e}}  \& {Finlator}}{{Mitra}
  et~al.}{2015}]{Mitra+2015}
{Mitra} S.,  {Dav{\'e}} R.,   {Finlator} K.,  2015, \mn@doi [\mnras]
  {10.1093/mnras/stv1387}, \href
  {http://adsabs.harvard.edu/abs/2015MNRAS.452.1184M} {452, 1184}

\bibitem[\protect\citeauthoryear{{Mobasher} et~al.,}{{Mobasher}
  et~al.}{2015}]{Mobasher+2015}
{Mobasher} B.,  et~al., 2015, \mn@doi [\apj] {10.1088/0004-637X/808/1/101},
  \href {https://ui.adsabs.harvard.edu/abs/2015ApJ...808..101M} {808, 101}

\bibitem[\protect\citeauthoryear{{More}, {van den Bosch}, {Cacciato}, {Skibba},
  {Mo}  \& {Yang}}{{More} et~al.}{2011}]{More+2011}
{More} S.,  {van den Bosch} F.~C.,  {Cacciato} M.,  {Skibba} R.,  {Mo} H.~J.,
  {Yang} X.,  2011, \mn@doi [\mnras] {10.1111/j.1365-2966.2010.17436.x}, \href
  {http://adsabs.harvard.edu/abs/2011MNRAS.410..210M} {410, 210}

\bibitem[\protect\citeauthoryear{{Moster}, {Naab}  \& {White}}{{Moster}
  et~al.}{2013}]{Moster+2013}
{Moster} B.~P.,  {Naab} T.,   {White} S.~D.~M.,  2013, \mn@doi [\mnras]
  {10.1093/mnras/sts261}, \href
  {http://adsabs.harvard.edu/abs/2013MNRAS.428.3121M} {428, 3121}

\bibitem[\protect\citeauthoryear{{Moster}, {Naab}  \& {White}}{{Moster}
  et~al.}{2018}]{Moster+2018}
{Moster} B.~P.,  {Naab} T.,   {White} S. D.~M.,  2018, \mn@doi [\mnras]
  {10.1093/mnras/sty655}, \href
  {https://ui.adsabs.harvard.edu/abs/2018MNRAS.477.1822M} {477, 1822}

\bibitem[\protect\citeauthoryear{{Moustakas} et~al.,}{{Moustakas}
  et~al.}{2013}]{Moustakas+2013}
{Moustakas} J.,  et~al., 2013, \mn@doi [\apj] {10.1088/0004-637X/767/1/50},
  \href {http://adsabs.harvard.edu/abs/2013ApJ...767...50M} {767, 50}

\bibitem[\protect\citeauthoryear{{Murray}, {Power}  \& {Robotham}}{{Murray}
  et~al.}{2013}]{Murray+2013}
{Murray} S.~G.,  {Power} C.,   {Robotham} A.~S.~G.,  2013, \mn@doi [Astronomy
  and Computing] {10.1016/j.ascom.2013.11.001}, \href
  {https://ui.adsabs.harvard.edu/abs/2013A&C.....3...23M} {3, 23}

\bibitem[\protect\citeauthoryear{{Napolitano} et~al.,}{{Napolitano}
  et~al.}{2024}]{Napolitano+2024}
{Napolitano} L.,  et~al., 2024, arXiv e-prints, \href
  {https://ui.adsabs.harvard.edu/abs/2024arXiv241010967N} {p. arXiv:2410.10967}

\bibitem[\protect\citeauthoryear{{Navarro-Carrera}, {Rinaldi}, {Caputi},
  {Iani}, {Kokorev}  \& {van Mierlo}}{{Navarro-Carrera}
  et~al.}{2024}]{Navarro-Carrera+2024}
{Navarro-Carrera} R.,  {Rinaldi} P.,  {Caputi} K.~I.,  {Iani} E.,  {Kokorev}
  V.,   {van Mierlo} S.~E.,  2024, \mn@doi [\apj] {10.3847/1538-4357/ad0df6},
  \href {https://ui.adsabs.harvard.edu/abs/2024ApJ...961..207N} {961, 207}

\bibitem[\protect\citeauthoryear{{Neistein}, {Li}, {Khochfar}, {Weinmann},
  {Shankar}  \& {Boylan-Kolchin}}{{Neistein} et~al.}{2011}]{Neistein+2011}
{Neistein} E.,  {Li} C.,  {Khochfar} S.,  {Weinmann} S.~M.,  {Shankar} F.,
  {Boylan-Kolchin} M.,  2011, \mn@doi [\mnras]
  {10.1111/j.1365-2966.2011.19145.x}, \href
  {http://adsabs.harvard.edu/abs/2011MNRAS.416.1486N} {416, 1486}

\bibitem[\protect\citeauthoryear{{Neto} et~al.,}{{Neto}
  et~al.}{2007}]{Neto+2007}
{Neto} A.~F.,  et~al., 2007, \mn@doi [\mnras]
  {10.1111/j.1365-2966.2007.12381.x}, \href
  {https://ui.adsabs.harvard.edu/abs/2007MNRAS.381.1450N} {381, 1450}

\bibitem[\protect\citeauthoryear{{Niemiec}, {Jullo}, {Giocoli}, {Limousin}  \&
  {Jauzac}}{{Niemiec} et~al.}{2019}]{Niemiec+2019}
{Niemiec} A.,  {Jullo} E.,  {Giocoli} C.,  {Limousin} M.,   {Jauzac} M.,  2019,
  \mn@doi [\mnras] {10.1093/mnras/stz1318}, \href
  {https://ui.adsabs.harvard.edu/abs/2019MNRAS.487..653N} {487, 653}

\bibitem[\protect\citeauthoryear{{Patel} et~al.,}{{Patel}
  et~al.}{2013}]{Patel+2013}
{Patel} S.~G.,  et~al., 2013, \mn@doi [\apj] {10.1088/0004-637X/766/1/15},
  \href {http://adsabs.harvard.edu/abs/2013ApJ...766...15P} {766, 15}

\bibitem[\protect\citeauthoryear{{Peng} et~al.,}{{Peng}
  et~al.}{2010}]{Peng+2010}
{Peng} Y.-j.,  et~al., 2010, \mn@doi [\apj] {10.1088/0004-637X/721/1/193},
  \href {http://adsabs.harvard.edu/abs/2010ApJ...721..193P} {721, 193}

\bibitem[\protect\citeauthoryear{{Pillepich} et~al.,}{{Pillepich}
  et~al.}{2018}]{Pillepich+2018}
{Pillepich} A.,  et~al., 2018, \mn@doi [\mnras] {10.1093/mnras/stx2656}, \href
  {http://adsabs.harvard.edu/abs/2018MNRAS.473.4077P} {473, 4077}

\bibitem[\protect\citeauthoryear{{Planck Collaboration} et~al.,}{{Planck
  Collaboration} et~al.}{2016}]{Planck+2015}
{Planck Collaboration} et~al., 2016, \mn@doi [\aap]
  {10.1051/0004-6361/201525830}, \href
  {http://adsabs.harvard.edu/abs/2016A%26A...594A..13P} {594, A13}

\bibitem[\protect\citeauthoryear{{Porras-Valverde}, {Forbes}, {Somerville},
  {Stevens}, {Holley-Bockelmann}, {Berlind}  \& {Genel}}{{Porras-Valverde}
  et~al.}{2023}]{Porras-Valverde+2023}
{Porras-Valverde} A.~J.,  {Forbes} J.~C.,  {Somerville} R.~S.,  {Stevens} A.
  R.~H.,  {Holley-Bockelmann} K.,  {Berlind} A.~A.,   {Genel} S.,  2023,
  \mn@doi [arXiv e-prints] {10.48550/arXiv.2310.11507}, \href
  {https://ui.adsabs.harvard.edu/abs/2023arXiv231011507P} {p. arXiv:2310.11507}

\bibitem[\protect\citeauthoryear{{Primack}}{{Primack}}{2024}]{Primack2024}
{Primack} J.~R.,  2024, \mn@doi [Annual Review of Nuclear and Particle Science]
  {10.1146/annurev-nucl-102622-023052}, \href
  {https://ui.adsabs.harvard.edu/abs/2024ARNPS..74..173P} {74, 173}

\bibitem[\protect\citeauthoryear{{Reddick}, {Wechsler}, {Tinker}  \&
  {Behroozi}}{{Reddick} et~al.}{2013}]{Reddick+2013}
{Reddick} R.~M.,  {Wechsler} R.~H.,  {Tinker} J.~L.,   {Behroozi} P.~S.,  2013,
  \mn@doi [\apj] {10.1088/0004-637X/771/1/30}, \href
  {http://adsabs.harvard.edu/abs/2013ApJ...771...30R} {771, 30}

\bibitem[\protect\citeauthoryear{{Rodriguez-Gomez} et~al.,}{{Rodriguez-Gomez}
  et~al.}{2016}]{Rodriguez-Gomez+2016}
{Rodriguez-Gomez} V.,  et~al., 2016, \mn@doi [\mnras] {10.1093/mnras/stw456},
  \href {http://adsabs.harvard.edu/abs/2016MNRAS.458.2371R} {458, 2371}

\bibitem[\protect\citeauthoryear{{Rodriguez-Puebla}}{{Rodriguez-Puebla}}{2024}]{Rodriguez-Puebla2024}
{Rodriguez-Puebla} A.,  2024, \mn@doi [arXiv e-prints]
  {10.48550/arXiv.2404.10801}, \href
  {https://ui.adsabs.harvard.edu/abs/2024arXiv240410801R} {p. arXiv:2404.10801}

\bibitem[\protect\citeauthoryear{{Rodr{\'\i}guez-Puebla}, {Drory}  \&
  {Avila-Reese}}{{Rodr{\'\i}guez-Puebla} et~al.}{2012}]{Rodriguez-Puebla+2012}
{Rodr{\'\i}guez-Puebla} A.,  {Drory} N.,   {Avila-Reese} V.,  2012, \mn@doi
  [\apj] {10.1088/0004-637X/756/1/2}, \href
  {https://ui.adsabs.harvard.edu/abs/2012ApJ...756....2R} {756, 2}

\bibitem[\protect\citeauthoryear{{Rodr{\'{\i}}guez-Puebla}, {Avila-Reese}  \&
  {Drory}}{{Rodr{\'{\i}}guez-Puebla} et~al.}{2013}]{Rodriguez-Puebla+2013}
{Rodr{\'{\i}}guez-Puebla} A.,  {Avila-Reese} V.,   {Drory} N.,  2013, \mn@doi
  [\apj] {10.1088/0004-637X/767/1/92}, \href
  {http://adsabs.harvard.edu/abs/2013ApJ...767...92R} {767, 92}

\bibitem[\protect\citeauthoryear{{Rodr{\'{\i}}guez-Puebla}, {Avila-Reese},
  {Yang}, {Foucaud}, {Drory}  \& {Jing}}{{Rodr{\'{\i}}guez-Puebla}
  et~al.}{2015}]{Rodriguez-Puebla+2015}
{Rodr{\'{\i}}guez-Puebla} A.,  {Avila-Reese} V.,  {Yang} X.,  {Foucaud} S.,
  {Drory} N.,   {Jing} Y.~P.,  2015, \mn@doi [\apj]
  {10.1088/0004-637X/799/2/130}, \href
  {http://adsabs.harvard.edu/abs/2015ApJ...799..130R} {799, 130}

\bibitem[\protect\citeauthoryear{{Rodr{\'{\i}}guez-Puebla}, {Primack},
  {Behroozi}  \& {Faber}}{{Rodr{\'{\i}}guez-Puebla}
  et~al.}{2016a}]{Rodriguez-Puebla+2016a}
{Rodr{\'{\i}}guez-Puebla} A.,  {Primack} J.~R.,  {Behroozi} P.,   {Faber}
  S.~M.,  2016a, \mn@doi [\mnras] {10.1093/mnras/stv2513}, \href
  {http://adsabs.harvard.edu/abs/2016MNRAS.455.2592R} {455, 2592}

\bibitem[\protect\citeauthoryear{{Rodr{\'{\i}}guez-Puebla}, {Behroozi},
  {Primack}, {Klypin}, {Lee}  \& {Hellinger}}{{Rodr{\'{\i}}guez-Puebla}
  et~al.}{2016b}]{Rodriguez-Puebla+2016}
{Rodr{\'{\i}}guez-Puebla} A.,  {Behroozi} P.,  {Primack} J.,  {Klypin} A.,
  {Lee} C.,   {Hellinger} D.,  2016b, \mn@doi [\mnras] {10.1093/mnras/stw1705},
  \href {http://adsabs.harvard.edu/abs/2016MNRAS.462..893R} {462, 893}

\bibitem[\protect\citeauthoryear{{Rodr{\'{\i}}guez-Puebla}, {Primack},
  {Avila-Reese}  \& {Faber}}{{Rodr{\'{\i}}guez-Puebla}
  et~al.}{2017}]{Rodriguez-Puebla+2017}
{Rodr{\'{\i}}guez-Puebla} A.,  {Primack} J.~R.,  {Avila-Reese} V.,   {Faber}
  S.~M.,  2017, \mn@doi [\mnras] {10.1093/mnras/stx1172}, \href
  {http://adsabs.harvard.edu/abs/2017MNRAS.470..651R} {470, 651}

\bibitem[\protect\citeauthoryear{{Rodr{\'\i}guez-Puebla}, {Calette},
  {Avila-Reese}, {Rodriguez-Gomez}  \&
  {Huertas-Company}}{{Rodr{\'\i}guez-Puebla}
  et~al.}{2020}]{Rodriguez-Puebla+2020}
{Rodr{\'\i}guez-Puebla} A.,  {Calette} A.~R.,  {Avila-Reese} V.,
  {Rodriguez-Gomez} V.,   {Huertas-Company} M.,  2020, \mn@doi [\pasa]
  {10.1017/pasa.2020.15}, \href
  {https://ui.adsabs.harvard.edu/abs/2020PASA...37...24R} {37, e024}

\bibitem[\protect\citeauthoryear{{Simha}, {Weinberg}, {Dav{\'e}}, {Fardal},
  {Katz}  \& {Oppenheimer}}{{Simha} et~al.}{2012}]{Simha+2012}
{Simha} V.,  {Weinberg} D.~H.,  {Dav{\'e}} R.,  {Fardal} M.,  {Katz} N.,
  {Oppenheimer} B.~D.,  2012, \mn@doi [\mnras]
  {10.1111/j.1365-2966.2012.21142.x}, \href
  {http://adsabs.harvard.edu/abs/2012MNRAS.423.3458S} {423, 3458}

\bibitem[\protect\citeauthoryear{{Somerville} \& {Dav{\'e}}}{{Somerville} \&
  {Dav{\'e}}}{2015}]{Somerville+2015}
{Somerville} R.~S.,  {Dav{\'e}} R.,  2015, \mn@doi [\araa]
  {10.1146/annurev-astro-082812-140951}, \href
  {http://adsabs.harvard.edu/abs/2015ARA%26A..53...51S} {53, 51}

\bibitem[\protect\citeauthoryear{{Stefanon} et~al.,}{{Stefanon}
  et~al.}{2015}]{Stefanon+2015}
{Stefanon} M.,  et~al., 2015, \mn@doi [\apj] {10.1088/0004-637X/803/1/11},
  \href {https://ui.adsabs.harvard.edu/abs/2015ApJ...803...11S} {803, 11}

\bibitem[\protect\citeauthoryear{{Steinhardt}, {Capak}, {Masters}  \&
  {Speagle}}{{Steinhardt} et~al.}{2016}]{Steinhardt+2016}
{Steinhardt} C.~L.,  {Capak} P.,  {Masters} D.,   {Speagle} J.~S.,  2016,
  \mn@doi [\apj] {10.3847/0004-637X/824/1/21}, \href
  {https://ui.adsabs.harvard.edu/abs/2016ApJ...824...21S} {824, 21}

\bibitem[\protect\citeauthoryear{{Tinker}, {Kravtsov}, {Klypin}, {Abazajian},
  {Warren}, {Yepes}, {Gottl{\"o}ber}  \& {Holz}}{{Tinker}
  et~al.}{2008}]{Tinker+2008}
{Tinker} J.,  {Kravtsov} A.~V.,  {Klypin} A.,  {Abazajian} K.,  {Warren} M.,
  {Yepes} G.,  {Gottl{\"o}ber} S.,   {Holz} D.~E.,  2008, \mn@doi [\apj]
  {10.1086/591439}, \href {http://adsabs.harvard.edu/abs/2008ApJ...688..709T}
  {688, 709}

\bibitem[\protect\citeauthoryear{{Vale} \& {Ostriker}}{{Vale} \&
  {Ostriker}}{2004a}]{ValeOstriker04}
{Vale} A.,  {Ostriker} J.~P.,  2004a, \mn@doi [\mnras]
  {10.1111/j.1365-2966.2004.08059.x}, \href
  {http://adsabs.harvard.edu/abs/2004MNRAS.353..189V} {353, 189}

\bibitem[\protect\citeauthoryear{{Vale} \& {Ostriker}}{{Vale} \&
  {Ostriker}}{2004b}]{ValeOstriker2004}
{Vale} A.,  {Ostriker} J.~P.,  2004b, \mn@doi [\mnras]
  {10.1111/j.1365-2966.2004.08059.x}, \href
  {http://adsabs.harvard.edu/abs/2004MNRAS.353..189V} {353, 189}

\bibitem[\protect\citeauthoryear{Virtanen et~al.,}{Virtanen
  et~al.}{2020}]{2020SciPy-NMeth}
Virtanen P.,  et~al., 2020, \mn@doi [Nature Methods]
  {10.1038/s41592-019-0686-2}, \href {https://rdcu.be/b08Wh} {17, 261}

\bibitem[\protect\citeauthoryear{{Wang}, {Mo}, {Li}  \& {Chen}}{{Wang}
  et~al.}{2023}]{Wang+2023}
{Wang} K.,  {Mo} H.,  {Li} C.,   {Chen} Y.,  2023, \mn@doi [\mnras]
  {10.1093/mnras/stad262}, \href
  {https://ui.adsabs.harvard.edu/abs/2023MNRAS.520.1774W} {520, 1774}

\bibitem[\protect\citeauthoryear{{Watson} et~al.,}{{Watson}
  et~al.}{2015}]{Watson15}
{Watson} D.~F.,  et~al., 2015, \mn@doi [\mnras] {10.1093/mnras/stu2065}, \href
  {http://adsabs.harvard.edu/abs/2015MNRAS.446..651W} {446, 651}

\bibitem[\protect\citeauthoryear{{Weaver} et~al.,}{{Weaver}
  et~al.}{2023}]{Weaver+2023}
{Weaver} J.~R.,  et~al., 2023, \mn@doi [\aap] {10.1051/0004-6361/202245581},
  \href {https://ui.adsabs.harvard.edu/abs/2023A&A...677A.184W} {677, A184}

\bibitem[\protect\citeauthoryear{{Wechsler} \& {Tinker}}{{Wechsler} \&
  {Tinker}}{2018}]{Wechsler+2018}
{Wechsler} R.~H.,  {Tinker} J.~L.,  2018, \mn@doi [\araa]
  {10.1146/annurev-astro-081817-051756}, \href
  {https://ui.adsabs.harvard.edu/abs/2018ARA&A..56..435W} {56, 435}

\bibitem[\protect\citeauthoryear{{Wechsler}, {Bullock}, {Primack}, {Kravtsov}
  \& {Dekel}}{{Wechsler} et~al.}{2002}]{Wechsler+2002}
{Wechsler} R.~H.,  {Bullock} J.~S.,  {Primack} J.~R.,  {Kravtsov} A.~V.,
  {Dekel} A.,  2002, \mn@doi [\apj] {10.1086/338765}, \href
  {http://adsabs.harvard.edu/abs/2002ApJ...568...52W} {568, 52}

\bibitem[\protect\citeauthoryear{{Weibel} et~al.,}{{Weibel}
  et~al.}{2024}]{Weibel+2024}
{Weibel} A.,  et~al., 2024, \mn@doi [\mnras] {10.1093/mnras/stae1891}, \href
  {https://ui.adsabs.harvard.edu/abs/2024MNRAS.tmp.1880W} {}

\bibitem[\protect\citeauthoryear{{Wetzel} \& {White}}{{Wetzel} \&
  {White}}{2010}]{Wetzel+2010}
{Wetzel} A.~R.,  {White} M.,  2010, \mn@doi [\mnras]
  {10.1111/j.1365-2966.2009.16191.x}, \href
  {http://adsabs.harvard.edu/abs/2010MNRAS.403.1072W} {403, 1072}

\bibitem[\protect\citeauthoryear{{Xiao} et~al.,}{{Xiao}
  et~al.}{2023}]{Xiao+2023}
{Xiao} M.,  et~al., 2023, \mn@doi [arXiv e-prints] {10.48550/arXiv.2309.02492},
  \href {https://ui.adsabs.harvard.edu/abs/2023arXiv230902492X} {p.
  arXiv:2309.02492}

\bibitem[\protect\citeauthoryear{{Yang}, {Mo}, {Zhang}  \& {van den
  Bosch}}{{Yang} et~al.}{2011}]{Yang+2011}
{Yang} X.,  {Mo} H.~J.,  {Zhang} Y.,   {van den Bosch} F.~C.,  2011, \mn@doi
  [\apj] {10.1088/0004-637X/741/1/13}, \href
  {https://ui.adsabs.harvard.edu/abs/2011ApJ...741...13Y} {741, 13}

\bibitem[\protect\citeauthoryear{{Yang}, {Mo}, {van den Bosch}, {Zhang}  \&
  {Han}}{{Yang} et~al.}{2012}]{Yang+2012}
{Yang} X.,  {Mo} H.~J.,  {van den Bosch} F.~C.,  {Zhang} Y.,   {Han} J.,  2012,
  \mn@doi [\apj] {10.1088/0004-637X/752/1/41}, \href
  {http://adsabs.harvard.edu/abs/2012ApJ...752...41Y} {752, 41}

\bibitem[\protect\citeauthoryear{{van Dokkum} et~al.,}{{van Dokkum}
  et~al.}{2010}]{vanDokkum+2010}
{van Dokkum} P.~G.,  et~al., 2010, \mn@doi [\apj]
  {10.1088/0004-637X/709/2/1018}, \href
  {http://adsabs.harvard.edu/abs/2010ApJ...709.1018V} {709, 1018}

\bibitem[\protect\citeauthoryear{{van de Voort}}{{van de
  Voort}}{2016}]{vandeVoort_2016}
{van de Voort} F.,  2016, \mn@doi [\mnras] {10.1093/mnras/stw1690}, \href
  {https://ui.adsabs.harvard.edu/abs/2016MNRAS.462..778V} {462, 778}

\bibitem[\protect\citeauthoryear{{van den Bosch}, {Jiang}, {Hearin},
  {Campbell}, {Watson}  \& {Padmanabhan}}{{van den Bosch}
  et~al.}{2014}]{vandenBosch+2014}
{van den Bosch} F.~C.,  {Jiang} F.,  {Hearin} A.,  {Campbell} D.,  {Watson} D.,
    {Padmanabhan} N.,  2014, \mn@doi [\mnras] {10.1093/mnras/stu1872}, \href
  {https://ui.adsabs.harvard.edu/abs/2014MNRAS.445.1713V} {445, 1713}

\makeatother
\end{thebibliography}


\bsp	
\label{lastpage}
\end{document}